%% file: access.tex
\let\TeXyear\year
\documentclass{ieeeaccess}

\let\year\TeXyear

\usepackage{cite}
\usepackage{amsmath,amssymb,amsfonts}
\usepackage{algorithmic}
\usepackage{graphicx}
\usepackage{textcomp}
\usepackage{xcolor}

\usepackage{soul}
\usepackage{todonotes}
\usepackage{dblfloatfix} 
\makeatletter
\let\MYcaption\@makecaption
\makeatother
\usepackage[font=footnotesize,labelformat=simple]{subcaption}

\makeatletter
\let\@makecaption\MYcaption
\makeatother
\usepackage{tikz}
\usepackage{pgfplots}
\usetikzlibrary{shapes,arrows,positioning,calc, matrix,spy,patterns}

\NewSpotColorSpace{PANTONE}
\AddSpotColor{PANTONE} {PANTONE3015C} {PANTONE\SpotSpace 3015\SpotSpace C} {1 0.3 0 0.2}
\SetPageColorSpace{PANTONE}
\definecolor{accessblue}{cmyk}{1, 0.3, 0, 0.2}
\definecolor{greycolor}{cmyk}{0,0,0,.8}

\usepackage[nolist]{acronym}
\pgfplotsset{compat=1.18}

\usepackage{booktabs}
\usepackage{xfrac}
\usepackage{comment}
\usepackage{orcidlink}

\definecolor{cb-1}{HTML}{4477AA}
\definecolor{cb-2}{HTML}{EE6677}
\definecolor{cb-3}{HTML}{228833}
\definecolor{cb-4}{HTML}{CCBB44}
\definecolor{cb-5}{HTML}{66CCEE}
\definecolor{cb-6}{HTML}{AA3377}
\definecolor{cb-7}{HTML}{BBBBBB}

\pgfplotscreateplotcyclelist{cb list}{
	cb-1,cb-2,cb-3,cb-4,cb-5,cb-6,cb-7
}

\makeatletter
\AtBeginDocument{\DeclareMathVersion{bold}
\SetSymbolFont{operators}{bold}{T1}{times}{b}{n}
\SetSymbolFont{NewLetters}{bold}{T1}{times}{b}{it}
\SetMathAlphabet{\mathrm}{bold}{T1}{times}{b}{n}
\SetMathAlphabet{\mathit}{bold}{T1}{times}{b}{it}
\SetMathAlphabet{\mathbf}{bold}{T1}{times}{b}{n}
\SetMathAlphabet{\mathtt}{bold}{OT1}{pcr}{b}{n}
\SetSymbolFont{symbols}{bold}{OMS}{cmsy}{b}{n}
\renewcommand\boldmath{\@nomath\boldmath\mathversion{bold}}}
\makeatother

\input{makros/KIT_colors.tex}

\input{makros/math.tex}
\input{makros/tikz_makros.tex}

\def\BibTeX{{\rm B\kern-.05em{\sc i\kern-.025em b}\kern-.08em
    T\kern-.1667em\lower.7ex\hbox{E}\kern-.125emX}}

\begin{document}
\input{acronyms.tex}
\history{Received 21 July 2025, accepted 1 September 2025, date of publication 9 September 2025, date of current version 24 September 2025.}
\doi{10.1109/ACCESS.2025.3607561}

\title{Neural Network-based Single-carrier\\ Joint Communication and Sensing: Loss Design, Constellation Shaping and Precoding}
\author{\uppercase{Charlotte Muth}\authorrefmark{1}, \IEEEmembership{Graduate Student Member, IEEE},
\uppercase{Benedikt Geiger}\authorrefmark{1}, \IEEEmembership{Graduate Student Member, IEEE}, 
\uppercase{Daniel Gil Gaviria}\authorrefmark{1}, \IEEEmembership{Graduate Student Member, IEEE},
and \uppercase{Laurent Schmalen}\authorrefmark{1},
\IEEEmembership{Fellow, IEEE}}

\address[1]{Communications Engineering Lab (CEL), Karlsruhe Institute of Technology (KIT), 
		Hertzstr. 16, 76187 Karlsruhe, Germany}
\tfootnote{This work has received funding 
	from the German
	Federal Ministry of Education and Research (BMBF) within the projects
	Open6GHub (grant agreement 16KISK010) and KOMSENS-6G (grant agreement 16KISK123).\\
Parts of this paper have been presented at the International Workshop on Smart Antennas (WSA) 2024.~\cite{Muth2024}}

\markboth
{Muth \headeretal: NN-based Single-carrier JCAS: Loss Design, Constellation Shaping and Precoding}
{Muth \headeretal: NN-based Single-carrier JCAS: Loss Design, Constellation Shaping and Precoding}

\corresp{Corresponding author: Charlotte Muth (e-mail: muth@kit.edu).}

\begin{abstract}
We investigate the impact of higher-order modulation formats on the sensing performance of single-carrier \ac{JCAS} systems. Several separate components such as a beamformer, a modulator, a target detector, an \ac{AoA} estimator and a communication demapper are implemented as trainable \acp{NN}. We compare geometrically shaped modulation formats to a classical \ac{QAM} scheme. We assess the influence of multi-snapshot sensing and varying \ac{SNR} on the overall performance of the autoencoder-based system. To improve the training behavior of the system, we decouple the loss functions from the respective \ac{SNR} values and the number of sensing snapshots, using upper bounds of the sensing and communication performance, namely the \acl{CRB} for \ac{AoA} estimation and the mutual information for communication. The \ac{NN}-based sensing outperforms classical algorithms, such as a Neyman-Pearson based power detector for object detection and ESPRIT for \ac{AoA} estimation for both the trained constellations and \ac{QAM} at low \acp{SNR}. We show that the gap in sensing performance between classical and shaped modulation formats can be significantly reduced through multi-snapshot sensing.
Lastly, we demonstrate system extension to multi-user \acl{MIMO} to address the improvement of spatial efficiency when servicing multiple \aclp{UE}.
Our contribution emphasizes the importance of estimation bounds for training neural networks, especially when the trained solutions are deployed in varying \ac{SNR} conditions.
\end{abstract}


\titlepgskip=-21pt

\maketitle

\acresetall
\input{tex_files/content_nn_jcas_2.tex}

\begin{appendices}

\input{tex_files/attachment.tex}
\end{appendices}

\bibliography{IEEEabrv,literature_short.bib}
\bibliographystyle{IEEEtran}

\begin{IEEEbiography}[{\includegraphics[trim={0cm 1.1cm 0 1.2cm},width=1in,height=1.25in,clip,keepaspectratio]{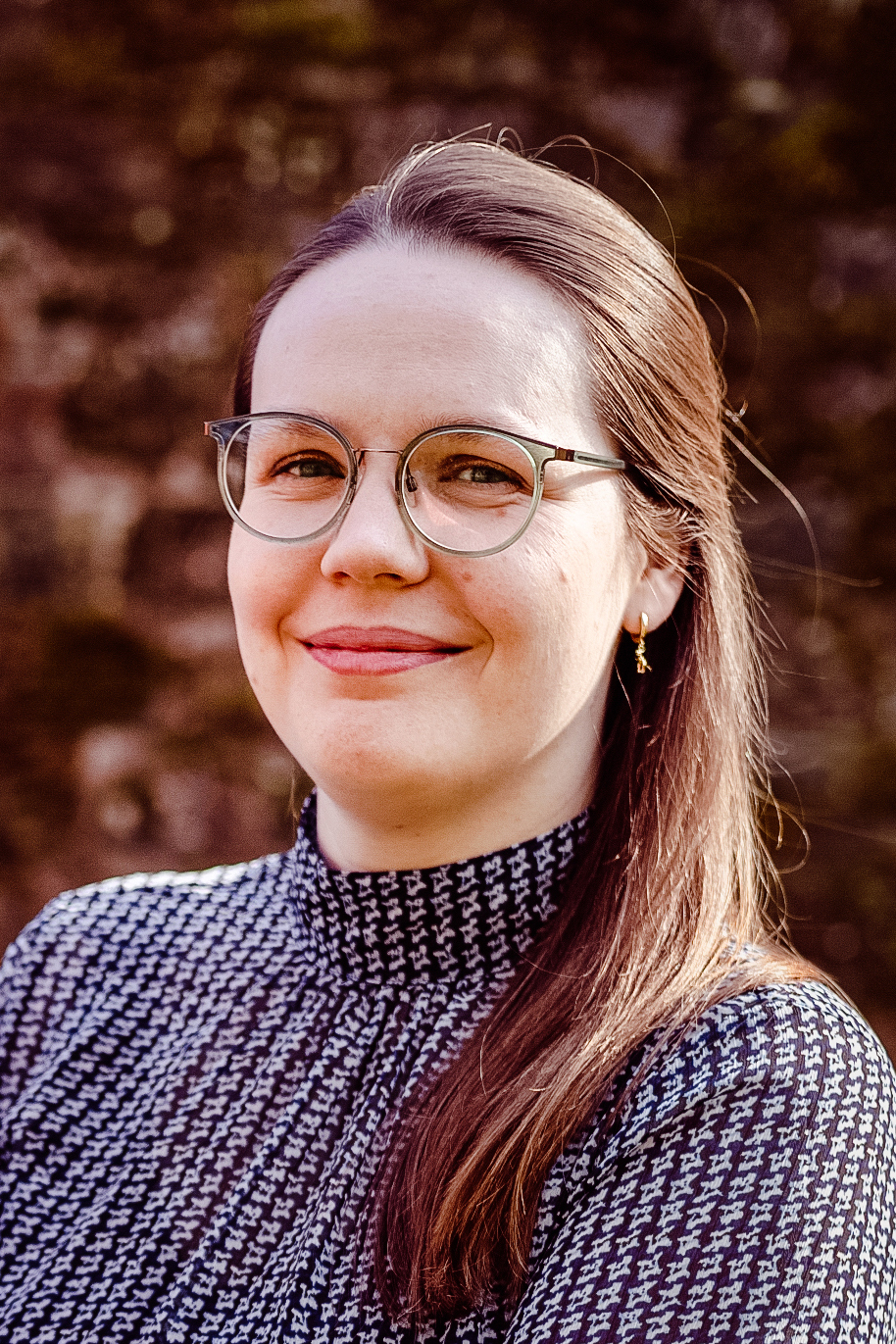}}]{Charlotte Muth} (Graduate Student Member, IEEE) received the B.Sc. and M.Sc. degrees in electrical engineering and information technology in 2019 and 2021, respectively from the Karlsruhe Institute of Technology, Karlsruhe, Germany. She is currently pursuing a Ph.D. at the Communications Engineering Lab at Karlsruhe Institute of Technology. 
Her research interests include the integration of sensing into communication systems, signal processing, and estimation using trainable algorithms and technologies for wireless communication. 
Ms. Muth has been part of the Executive Committee of the IEEE Student Branch at Karlsruhe Institute of Technology since 2022.
\end{IEEEbiography}
\vskip -2.5\baselineskip plus -1fil
\begin{IEEEbiography}[{\includegraphics[width=1in,height=1.25in,clip,keepaspectratio]{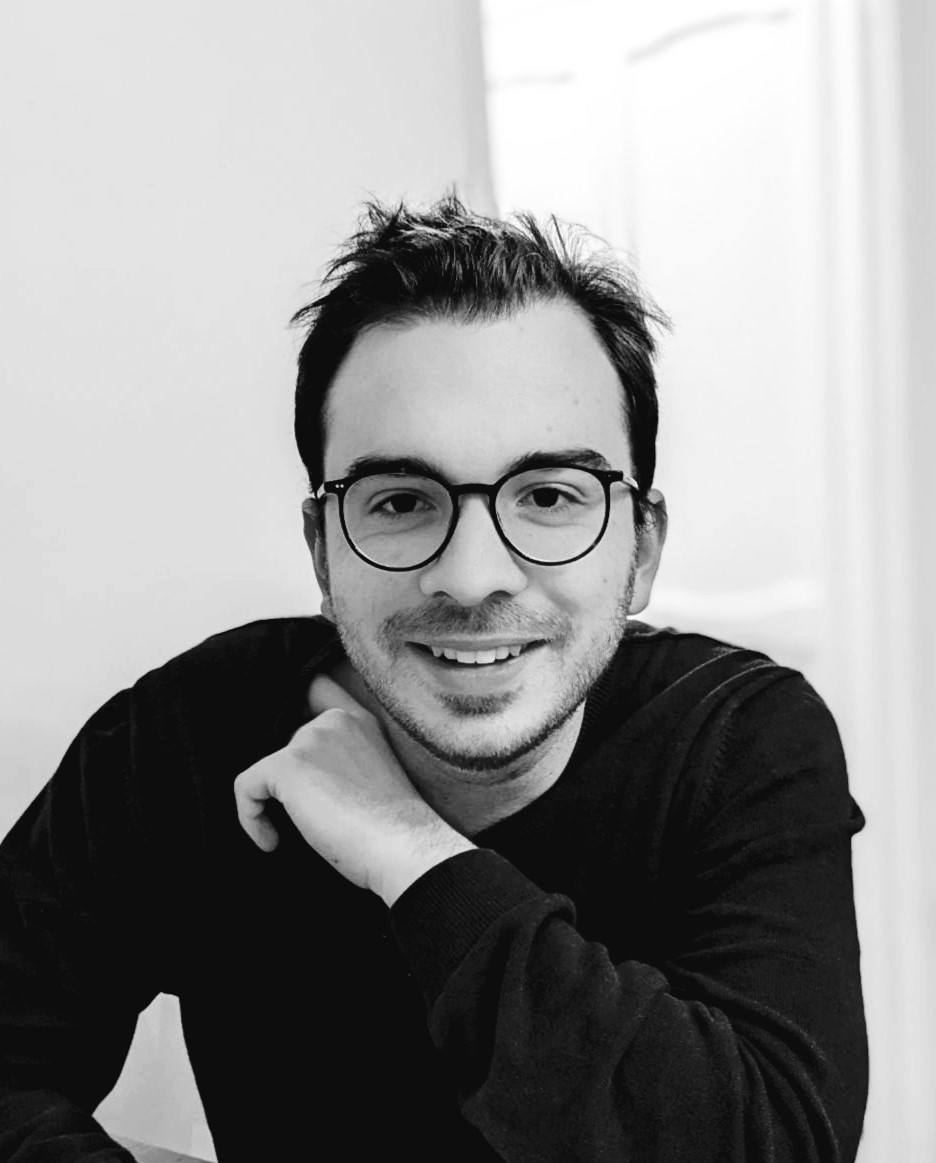}}]{Benedikt Geiger} (Graduate Student Member, IEEE) received the B.Sc. and M.Sc. degrees in electrical engineering and information technology from the Karlsruhe Institute of Technology (KIT), Karlsruhe, Germany, in 2020 and 2023, respectively, where he is currently working toward the Ph.D. degree. From 2021 to 2022, he was an ERASMUS+ Research Intern with the Optical Networks Group, University College London, London, U.K.. He wrote his master thesis with Nokia, Stuttgart, Germany, in 2022. His research interests include joint communication and sensing for 6G, ultra high-speed fiber optic communication systems, and the application of machine learning in communications. Mr. Geiger is a Student Member of the VDE, and his awards and honors include the first place of the Rohde \& Schwarz Engineering Competition 2021. During his studies, he was a Scholar of the Konrad Adenauer Foundation, and is an alumnus of the Karlsruhe Institute of Technology’s Leadership Talent Academy.
\end{IEEEbiography}
\vskip -2.5\baselineskip plus -1fil
\begin{IEEEbiography}[{\includegraphics[width=1in,height=1.25in,clip,keepaspectratio]{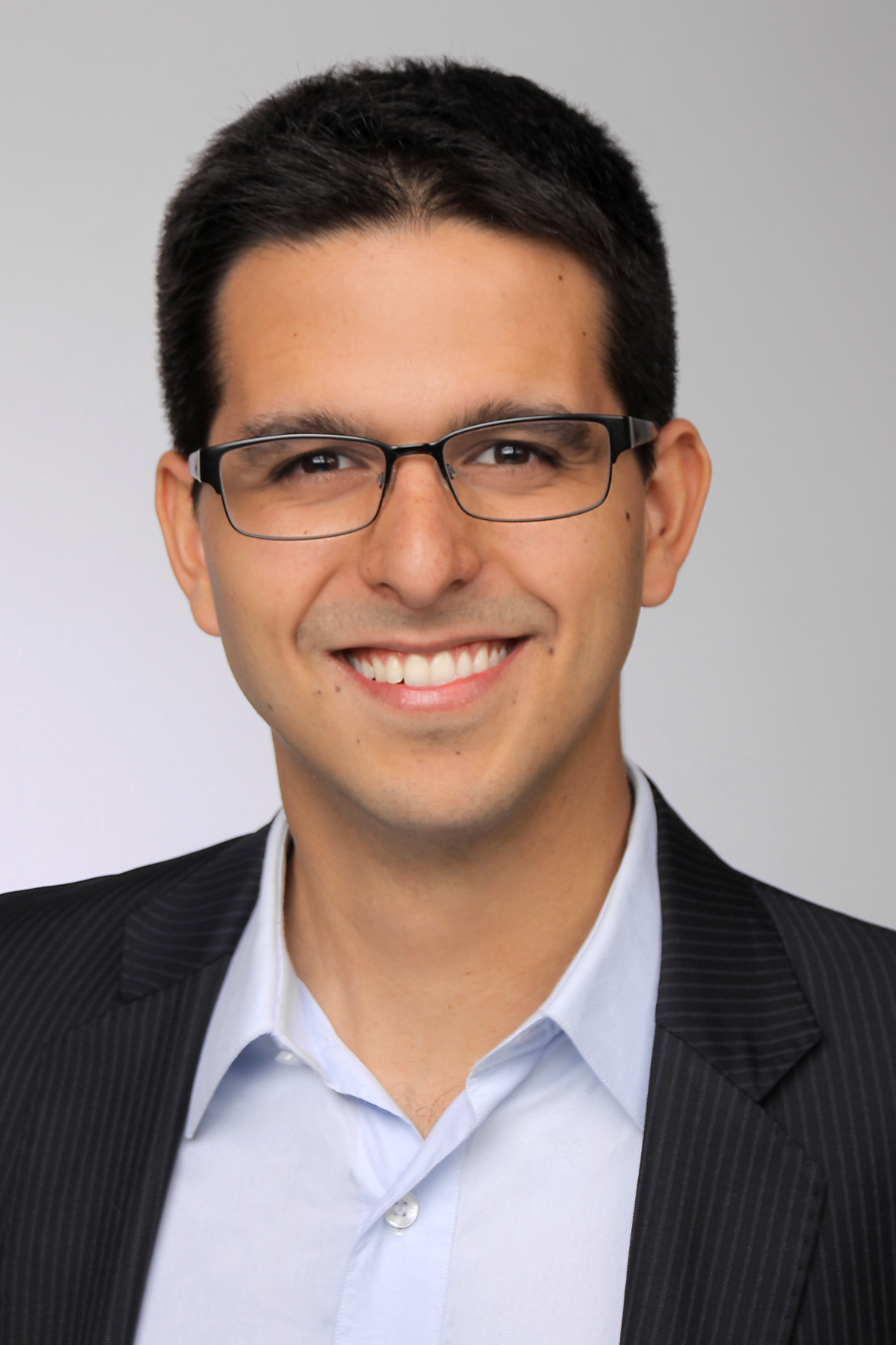}}]{Daniel Gil Gaviria} (Graduate Student Member,
IEEE) received the B.Sc. and M.Sc. degrees in
electrical engineering and information technology
from the Karlsruhe Institute of Technology (KIT),
Karlsruhe, Germany, in 2018 and 2020, respectively,
where he is currently pursuing the Ph.D. degree.
He is also a Research and Teaching Assistant
with the Communications Engineering Laboratory
(CEL), KIT. His research interests include modern
modulation schemes and signal processing for integrated sensing and communication (ISAC), specially in the context of automotive radar-communication and future cellular networks.
\end{IEEEbiography}
\vskip -2.5\baselineskip plus -1fil
\begin{IEEEbiography}[{\includegraphics[width=1in,height=1.25in,clip,keepaspectratio]{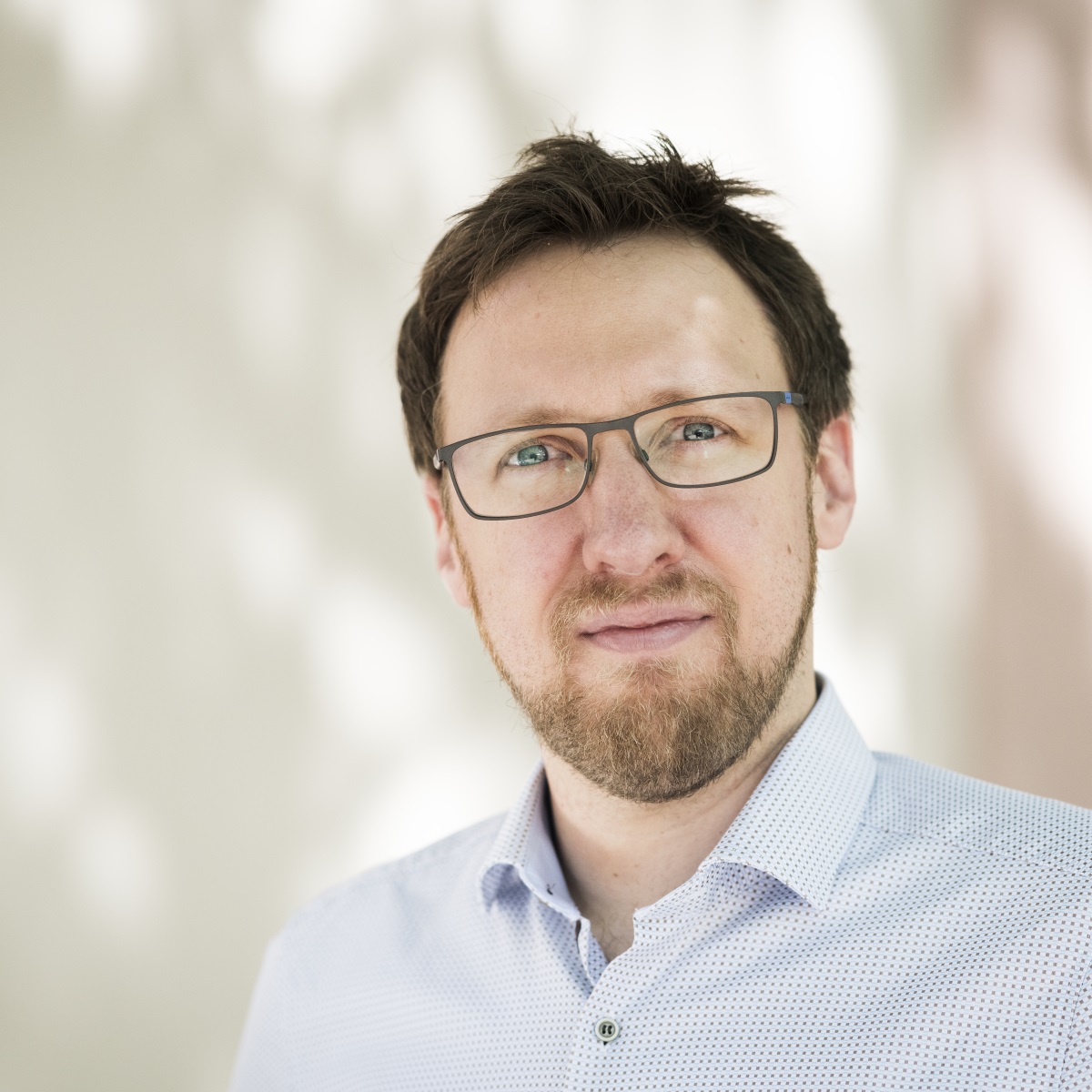}}]{Laurent Schmalen} (S'07--M'13--SM'16--F'23) received the Dipl.-Ing. and Dr.-Ing. degrees in electrical engineering and information technology from RWTH Aachen University, Germany, in 2005 and 2011, respectively. From 2011 to 2019, he was with Alcatel-Lucent Bell Labs and Nokia Bell Labs. From 2014 to 2019, he was a Guest Lecturer with the University of Stuttgart, Stuttgart, Germany. Since 2019, he has been a Full Professor with Karlsruhe Institute of Technology, where he co-heads the Communications Engineering Laboratory. His research has been funded, among others, by the European Research Council (ERC) via an ERC Consolidator Grant. His research interests include channel coding, modulation formats, and optical communications. He was a recipient and a co-recipient of several awards, including the E-Plus Award for the Ph.D. thesis, the 2013 Best Student Paper Award from the IEEE Signal Processing Systems Workshop, the 2016 and 2018 Journal of Lightwave Technology Best Paper Awards, and the Bell Labs President Award. He serves as an Associate Editor for IEEE Transactions on Communications. He has served as TPC member for most major conferences in the fields of communications and information theory.
\end{IEEEbiography}
\EOD
\end{document}

%% file: makros/KIT_colors.tex
\definecolor{kit-green100}{rgb}{0,.59,.51}
\definecolor{kit-green70}{rgb}{.3,.71,.65}
\definecolor{kit-green50}{rgb}{.50,.79,.75}
\definecolor{kit-green30}{rgb}{.69,.87,.85}
\definecolor{kit-green15}{rgb}{.85,.93,.93}
\definecolor{KITgreen}{rgb}{0,.59,.51}

\definecolor{KITpalegreen}{RGB}{130,190,60}
\colorlet{kit-maigreen100}{KITpalegreen}
\colorlet{kit-maigreen70}{KITpalegreen!70}
\colorlet{kit-maigreen50}{KITpalegreen!50}
\colorlet{kit-maigreen30}{KITpalegreen!30}
\colorlet{kit-maigreen15}{KITpalegreen!15}

\definecolor{KITblue}{rgb}{.27,.39,.66}
\definecolor{kit-blue100}{rgb}{.27,.39,.67}
\definecolor{kit-blue70}{rgb}{.49,.57,.76}
\definecolor{kit-blue50}{rgb}{.64,.69,.83}
\definecolor{kit-blue30}{rgb}{.78,.82,.9}
\definecolor{kit-blue15}{rgb}{.89,.91,.95}

\definecolor{KITyellow}{rgb}{.98,.89,0}
\definecolor{kit-yellow100}{cmyk}{0,.05,1,0}
\definecolor{kit-yellow70}{cmyk}{0,.035,.7,0}
\definecolor{kit-yellow50}{cmyk}{0,.025,.5,0}
\definecolor{kit-yellow30}{cmyk}{0,.015,.3,0}
\definecolor{kit-yellow15}{cmyk}{0,.0075,.15,0}

\definecolor{KITorange}{rgb}{.87,.60,.10}
\definecolor{kit-orange100}{cmyk}{0,.45,1,0}
\definecolor{kit-orange70}{cmyk}{0,.315,.7,0}
\definecolor{kit-orange50}{cmyk}{0,.225,.5,0}
\definecolor{kit-orange30}{cmyk}{0,.135,.3,0}
\definecolor{kit-orange15}{cmyk}{0,.0675,.15,0}

\definecolor{KITred}{rgb}{.63,.13,.13}
\definecolor{kit-red100}{cmyk}{.25,1,1,0}
\definecolor{kit-red70}{cmyk}{.175,.7,.7,0}
\definecolor{kit-red50}{cmyk}{.125,.5,.5,0}
\definecolor{kit-red30}{cmyk}{.075,.3,.3,0}
\definecolor{kit-red15}{cmyk}{.0375,.15,.15,0}

\definecolor{KITpurple}{RGB}{160,0,120}
\colorlet{kit-purple100}{KITpurple}
\colorlet{kit-purple70}{KITpurple!70}
\colorlet{kit-purple50}{KITpurple!50}
\colorlet{kit-purple30}{KITpurple!30}
\colorlet{kit-purple15}{KITpurple!15}

\definecolor{KITcyanblue}{RGB}{80,170,230}
\colorlet{kit-cyanblue100}{KITcyanblue}
\colorlet{kit-cyanblue70}{KITcyanblue!70}
\colorlet{kit-cyanblue50}{KITcyanblue!50}
\colorlet{kit-cyanblue30}{KITcyanblue!30}
\colorlet{kit-cyanblue15}{KITcyanblue!15}

\definecolor{cb-1}{HTML}{4477AA}
\definecolor{cb-2}{HTML}{EE6677}
\definecolor{cb-3}{HTML}{228833}
\definecolor{cb-4}{HTML}{CCBB44}
\definecolor{cb-5}{HTML}{66CCEE}
\definecolor{cb-6}{HTML}{AA3377}
\definecolor{cb-7}{HTML}{BBBBBB}

\pgfplotscreateplotcyclelist{cb list}{
	cb-1,cb-2,cb-3,cb-4,cb-5,cb-6,cb-7
}

%% file: makros/math.tex
\let\Re\relax
\let\Im\relax
\DeclareMathOperator{\Re}{Re}
\DeclareMathOperator{\Im}{Im}

\DeclareMathAlphabet{\mathbfsf}{\encodingdefault}{\sfdefault}{bx}{n}

\newcommand{\mvrv}[1]{\boldsymbol{\MakeLowercase{\mathsf{#1}}}}

\newcommand*{\vect}[1]{\boldsymbol{#1}}
\newcommand*{\mat}[1]{\MakeUppercase{\boldsymbol{#1}}}

%% file: makros/tikz_makros.tex
\usepackage{ellipsis}
\usetikzlibrary{calc, patterns,fit,shapes}
\usetikzlibrary{decorations.pathreplacing,decorations.markings,shapes.geometric}
\usepgfplotslibrary{fillbetween}
\tikzset{naming/.style={align=center,font=\small}}
\tikzset{antenna/.style={insert path={-- coordinate (ant#1) ++(0,0.25) -- +(135:0.25) + (0,0) -- +(45:0.25)}}}
\tikzset{station/.style={naming,draw,shape=dart,shape border rotate=90, minimum width=10mm, minimum height=10mm,outer sep=0pt,inner sep=3pt}}
\tikzset{mobile/.style={naming,draw,shape=rectangle,minimum width=12mm,minimum height=6mm, outer sep=0pt,inner sep=3pt}}
\tikzset{radiation/.style={{decorate,decoration={expanding waves,angle=90,segment length=4pt}}}}
\usetikzlibrary{decorations.pathreplacing}

\tikzset{pics/MUE/.style args={#1}
{code={
\node[mobile,label={[inner ysep=+.3333em]\dots}] (box){};
\draw ([xshift=.25cm] box.south west) circle (4pt)
      ([xshift=-.25cm]box.south east) circle (4pt);

\fill ([xshift=.25cm] box.south west) circle (1pt)
      ([xshift=-.25cm]box.south east) circle (1pt);

\draw ([xshift=.25cm] box.north west) [antenna=1];
\draw ([xshift=-.25cm]box.north east) [antenna=2];
}}}

\tikzset{pics/UE/.style args={#1}
{code={
\node[mobile] (box) {#1};

\draw ([xshift=.25cm] box.south west) circle (4pt)
      ([xshift=-.25cm]box.south east) circle (4pt);

\fill ([xshift=.25cm] box.south west) circle (1pt)
      ([xshift=-.25cm]box.south east) circle (1pt);

\draw (box.north) [antenna=1];
}}}

\tikzset{pics/MBS/.style args={#1}{code={
\node[station] (-base) {#1};
\draw[line join=bevel] (-base.100) -- (-base.80) -- (-base.110) -- (-base.70) -- (-base.north west) -- (-base.north east);
\draw[line join=bevel] (-base.100) -- (-base.70) (-base.110) -- (-base.north east);
\node (-a1) at ([xshift=.3cm,yshift=.3cm] -base.north) {};
\node (-a2) at ([xshift=-.3cm,yshift=.3cm] -base.north) {};
\draw[line cap=rect] ([xshift=.3cm,yshift=.3pt] -base.north) [antenna=1];
\draw[line cap=rect] ([yshift=.3pt]ant1 |- -base.north) -- ([xshift=-.3cm,yshift=.3pt] -base.north) [antenna=2];
\node (-d) at ([yshift=1cm] -base.east) {\dots};

}}}

\tikzset{pics/BS/.style args={#1}{code={

\node[station] (base) {#1};

\draw[line join=bevel] (base.100) -- (base.80) -- (base.110) -- (base.70) -- (base.north west) -- (base.north east);
\draw[line join=bevel] (base.100) -- (base.70) (base.110) -- (base.north east);

\draw[line cap=rect] ([yshift=0pt]base.north) [antenna=1];

}}}

\tikzset{pics/car/.style args={#1}{code={
    \node (-base) at (4.2,2){};
    \draw[very thick,#1, rounded corners=0.18ex,fill=black!20!blue!20!white]  (2.7,1.8) -- ++(1,0.7) -- ++(1.6,0) -- ++(0.6,-0.7) -- (2.7,1.8);
        \draw[thick,#1, fill=#1,thick] (4.2,2.5) -- (5.3,2.5)  -- ++(0.2,-0.23) -- (4.4,2.27) -- (4.4,1.8) --(4.2,1.8) -- cycle;
      \draw[thick]  (4.2,1.8) -- (4.2,2.5);
      \fill[draw=#1,fill=#1,rounded corners=0.6ex,very thick] (1.5,.5) -- ++(0,1) -- ++(1.2,0.3) --  ++(3,0) -- ++(1,0) -- ++(0,-1.3) -- (1.5,.5) -- cycle;
      \draw[draw=black,fill=gray!50,thick] (2.75,.5) circle (.5);
      \draw[draw=black,fill=gray!50,thick] (5.5,.5) circle (.5);
      \draw[draw=black,fill=gray!80,semithick] (2.75,.5) circle (.4);
      \draw[draw=black,fill=gray!80,semithick] (5.5,.5) circle (.4);
      \filldraw[draw=#1, fill=KITorange, semithick] (2.1,1.3) -- ++(150:.5) arc (155:205:0.5) -- cycle;
  
}}}

\tikzset{
  pobl/.style={
    inner sep=0pt, outer sep=0pt, fill=#1,
  },
  pobl gron/.style n args={2}{
    pobl=#1, rounded corners=#2,
  },
  pics/person/.style n args={3}{
    code={
      \node (-corff) [pobl=#1, minimum width=.25*#2, minimum height=.375*#2, rotate=#3, pic actions] {};
      \node (-pen) [minimum width=.3*#2, circle, pobl=#1, outer sep=.01*#2, anchor=south, rotate=#3, pic actions] at (-corff.north) {};
      \node (-coes dde) [pobl gron={#1}{1pt}, anchor=north west, minimum width=.12125*#2, minimum height=.25*#2, rotate=#3, pic actions] at (-corff.south west) {};
      \node [pobl=#1, anchor=north, minimum width=.12125*#2, minimum height=.15*#2, rotate=#3, pic actions] at (-coes dde.north) {};
      \node (-coes chwith) [pobl gron={#1}{1pt}, anchor=north east, minimum width=.12125*#2, minimum height=.25*#2, rotate=#3, pic actions] at (-corff.south east) {};
      \node [pobl=#1, anchor=north, minimum width=.12125*#2, minimum height=.15*#2, rotate=#3, pic actions] at (-coes chwith.north) {};
      \node (-braich dde) [pobl gron={#1}{.75pt}, minimum width=.075*#2, minimum height=.325*#2, outer sep=.0064*#2, anchor=north west, rotate=#3, pic actions] at (-corff.north east)  {};
      \node [pobl=#1, minimum width=.05*#2, minimum height=.2*#2, outer sep=.0064*#2, anchor=north west, rotate=#3, pic actions] at (-corff.north east) {};
      \node (-braich chwith) [pobl gron={#1}{.75pt}, minimum width=.075*#2, minimum height=.325*#2, outer sep=.0064*#2, anchor=north east, rotate=#3, pic actions] at (-corff.north west) {};
      \node [pobl=#1, minimum width=.0375*#2, minimum height=.2*#2, outer sep=.0064*#2, anchor=north east, rotate=#3, pic actions] at (-corff.north west) {};
      \node (-fit person) [fit={(-pen.north) (-braich dde.east) (-coes chwith.south) (-braich chwith.west)}] {};
      \node (-pwy) [below=25pt of -fit person, every pin] {\tikzpictext};
      \draw [every pin edge] (-fit person) -- (-pwy);
    };
  };
}

\tikzset{pics/phone/.style args={#1}{code={
    \node (-base) at (1,1.75) {};
      \draw[draw=black,fill=black!80,rounded corners=0.2ex,very thick] (0,0) -- ++(0,3.5) -- ++(2,0) --  ++(0,-3.5) -- ++(-2,0) -- cycle;
      \fill[thick, fill=KITblue] (0.2,0.2) -- ++(0,3.1)  -- ++(1.6,0) -- ++(0,-3.1) -- ++(-1.6,0) -- cycle;
      \fill[fill=#1] (0.65,2) rectangle (0.75,2.1);
      \fill[fill=#1] (0.85,2) rectangle (0.95,2.3);
      \fill[fill=#1] (1.05,2) rectangle (1.15,2.5);
      \fill[fill=#1] (1.25,2) rectangle (1.35,2.7);
  
}}}

\newcommand{\hi}[2][KITyellow]{
	\ifmmode
	{
		 \mathchoice%
			{\colorbox{#1!50}{\textcolor{black}{$\displaystyle#2$}}}%
			{\colorbox{#1!50}{\textcolor{black}{$\textstyle#2$}}}
			{\colorbox{#1!50}{\textcolor{black}{$\scriptstyle#2$}}}
			{\colorbox{#1!50}{\textcolor{black}{$\scriptscriptstyle#2$}}}
	}
	\else
	{
		{\colorbox{#1!50}{\textcolor{black}{#2}}}
	}
	\fi
	}

%% file: acronyms.tex
\begin{acronym}[TROLOLO]
  \acro{ACM}{auto-correlation matrix}
  \acro{ADC}{analog to digital converter}
  \acro{AE}{autoencoder}
  \acro{ASK}{amplitude shift keying}
  \acro{APSK}{amplitude phase shift keying}
  \acro{AoA}{angle of arrival}
  \acro{AWGN}{additive white Gaussian noise}
  \acro{BER}{bit error rate}
  \acro{BCE}{binary cross entropy}
  \acro{BICM}{bit-interleaved coded modulation}
  \acro{BMI}{bit-wise mutual information}
  \acro{BPSK}{binary phase shift keying}
  \acro{BP}{backpropagation}
  \acro{BSC}{binary symmetric channel}
  \acro{CAZAC}{constant amplitude zero autocorrelation waveform}
  \acro{CDF}{cumulative distribution function}
  \acro{CE}{cross entropy}
  \acro{CNN}{concolutional neural network}
  \acro{CP}{cyclic prefix}
  \acro{CRB}{Cramér-Rao bound}
  \acro{CRC}{cyclic redundancy check}
  \acro{CSI}{channel state information}
  \acro{DFT}{discrete Fourier transform}
  \acro{DNN}{deep neural network}
  \acro{DoA}{degree of arrival}
  \acro{DOCSIS}{data over cable services}
  \acro{DPSK}{differential phase shift keying}
  \acro{DSL}{digital subscriber line}
  \acro{DSP}{digital signal processing}
  \acro{DTFT}{discrete-time Fourier transform}
  \acro{DVB}{digital video broadcasting}
  \acro{ELU}{exponential linear unit}
  \acro{ESPRIT}{Estimation of Signal Parameter via Rotational Invariance Techniques}
  \acro{FEC}{forward error correction}
  \acro{FFNN}{feed-forward neural network}
  \acro{FFT}{fast Fourier transform}
  \acro{FIR}{finite impulse response}
  \acro{GD}{gradient descent}
  \acro{GF}{Galois field}
  \acro{GMM}{Gaussian mixture model}
  \acro{GMI}{generalized mutual information}
  \acro{ICI}{inter-channel interference}
  \acro{IDE}{integrated development environment}
  \acro{IDFT}{inverse discrete Fourier transform}
  \acro{IFFT}{inverse fast Fourier transform}
  \acro{IIR}{infinite impulse response}
  \acro{ISI}{inter-symbol interference}
  \acro{JCAS}{joint communication and sensing}
  \acro{KKT}{Karush-Kuhn-Tucker}
  \acro{KLD}{Kullback-Leibler divergence}
  \acro{LDPC}{low-density parity-check}
  \acro{LLR}{log-likelihood ratio}
  \acro{LTE}{long-term evolution}
  \acro{LTI}{linear time-invariant}
  \acro{LR}{logistic regression}
  \acro{MAC}{multiply-accumulate}
  \acro{MAP}{maximum a posteriori}
  \acro{MIMO}{multiple-input multiple-output}
  \acro{MLP}{multilayer perceptron}
  \acro{MLD}{maximum likelihood demapper}
  \acro{ML}{machine learning}
  \acro{MSE}{mean squared error}
  \acro{MLSE}{maximum-likelihood sequence estimation}
  \acro{MMSE}{miminum mean squared error}
  \acro{NN}{neural network}
  \acro{NP}{Neyman-Pearson}
  \acro{OFDM}{orthogonal frequency-division multiplexing}
  \acro{OLA}{overlap-add}
  \acro{PAPR}{peak-to-average-power ratio}
  \acro{PDF}{probability density function}
  \acro{pmf}{probability mass function}
  \acro{PSD}{power spectral density}
  \acro{PSK}{phase shift keying}
  \acro{QAM}{quadrature amplitude modulation}
  \acro{QPSK}{quadrature phase shift keying}
  \acro{radar}{radio detection and ranging}
  \acro{RC}{raised cosine}
  \acro{RCS}{radar cross section}
  \acro{RMSE}{root mean squared error}
  \acro{RNN}{recurrent neural network}
  \acro{ROM}{read-only memory}
  \acro{RRC}{root raised cosine}
  \acro{RV}{random variable}
  \acro{SER}{symbol error rate}
  \acro{SNR}{signal-to-noise ratio}
  \acro{SINR}{signal-to-noise-and-interference ratio}
  \acro{SPA}{sum-product algorithm}
  \acro{UE}{user equipment}
  \acro{VCS}{version control system}
  \acro{WLAN}{wireless local area network}
  \acro{WSS}{wide-sense stationary}
\end{acronym}

%% file: tex_files/content_nn_jcas_2.tex
\acresetall
\section{Introduction}
Digital communication is a vital service for our hyper-connected society. The integration of a sensing service, which, e.g. aids automotive vehicles to detect potential collisions while serving communication users at a busy intersection, could improve the user experience of mobile network users. An increase in spectral and energy efficiency can be achieved by combining radio communication and sensing into one waveform instead of operating two separate systems. Therefore, this work focuses on the codesign of both functionalities in a \ac{JCAS} system.
The future 6G network is expected to natively support \ac{JCAS} by introducing object detection of objects without communication capabilities and by performing general sensing
of the surroundings \cite{Wild2021}. With this approach, we expect to increase the spectral efficiency by providing spectral resources for sensing while maintaining their use for communication, as well as the energy efficiency because of the dual use of a joint waveform.

There is growing interest in data-driven solutions based on \ac{ML} since they can overcome deficits such as hardware impairments, faced by algorithms derived using model-based techniques \cite{MateosRamos2021}. Especially at higher carrier frequencies, which will become more important in 6G, these deficits become more pronounced due to hardware imperfections \cite{MateosRamos2023}. \ac{ML} is expected to be prevalent in 6G, as its use in communication and radar signal processing has matured in recent years \cite{Wild2021}.

\acused{NN}
\begin{figure*}[t]
	\centerline{\input{figures/flowgraph_alternative}}
	\vspace{-0.4cm}
	\caption{\ac{JCAS} system, light blue blocks are trainable \acp{NN}. The modulator can be implemented as an \ac{NN} or with a classical \ac{QAM} constellation.}
	\label{fig:flowgraphtrain_mono}
	\vspace*{-0.3cm} 
\end{figure*}
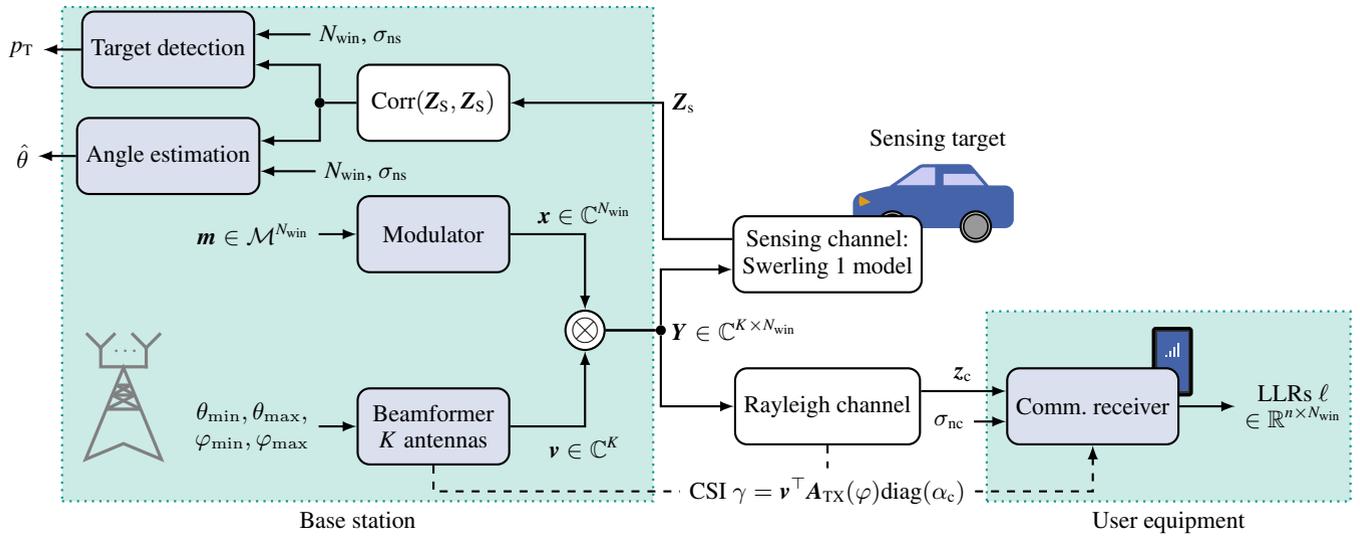
\makeatletter
\AC@reset{NN}
\makeatother
For a communication-centric \ac{JCAS} network, it is essential to first guarantee communication capabilities. Therefore, the effects of transmit signal design, such as the choice of modulation formats, are of particular interest. Many studies such as~\cite{Schieler2020,Oliveira2023} show sensing using only phase modulation, however, higher-order \ac{QAM} formats are often employed in legacy communication systems to increase throughput and reliability.

In this paper, we study the monostatic sensing capabilities of a wireless communication system with multiple snapshots. We consider single-carrier modulation to reduce the computational complexity of the simulations. Our findings can generally be transferred to multi-carrier systems such as \ac{OFDM}.
Our contributions are:
\begin{itemize}
    \item We compare an \ac{AE}-based geometric constellation shaping approach to classical \ac{QAM} to analyze in detail the impact on communication and sensing in different \ac{SNR} environments and multi-snapshot sensing.
    \item We modify the loss function for the end-to-end training to improve training over a wide range of \ac{SNR} conditions
    \item We show that the loss modification improves the \ac{AoA} estimation and that the detection and communication loss terms do not profit from modification.
    \item We extend the \ac{AE}-approach to demonstrate the servicing of multiple \acp{UE}.
\end{itemize}

The remainder of this paper is structured as follows. Section~\ref{sec:sysmodel} introduces the proposed system model. In Sec.~\ref{sec:NN}, the design and training of the \acp{DNN} are described. Section~\ref{sec:sims} introduces the benchmark algorithms which our system is compared to and presents the results of the simulations performed, including evaluation of \ac{BMI}, \ac{RMSE} of angle estimation, detection rate, and false alarm rate for object detection.
Lastly, in Sec.~\ref{sec:concl}, we conclude this work.
\subsection{Related Work}\label{sec:related-work}
\ac{ML} approaches allow the joint optimization of the transmitter and receiver taking into account sensing and communications performance metrics. \acp{AE} have been studied as a promising \ac{ML} approach for communication systems \cite{OShea2017,Cammerer2020}, and for radar \cite{JaraboAmores2008, Fuchs2020} independently of each other. As \ac{JCAS} has gained significant attention within the scientific community, \ac{ML} research for \ac{JCAS} has emerged, merging approaches for radar and communications.
In \cite{MateosRamos2021}, an \ac{AE} for \ac{JCAS} in a single-carrier system has been proposed, performing close to a maximum a posteriori ratio test detector benchmark for single snapshot sensing of one radar target. In~\cite{Muth2023}, this approach was extended to multiple targets, addressing the problem of target matching.
The work of \cite{MateosRamos2023} extends these methods to an \ac{OFDM} waveform using model-based learning, which is a well-known technique for combining communication and radar \cite{Sturm2011, Braun2010}. These works demonstrate the potential of deep learning-based sensing to mitigate hardware mismatches. 

However, the research has been limited to single snapshot estimation. We presented an extension to multiple-snapshot sensing in~\cite{Muth2024}, while limiting ourselves to QAM modulation. In this work, we train a system that can generalize over a range of \ac{SNR} for both sensing and communication, demonstrate the effect of constellation shaping and give a detailed analysis of our approach.

\subsection{Notation} 
$\mathbb{R}$ and $\mathbb{C}$ denote the set of real and complex numbers, respectively. Sets are generally denoted by calligraphic font, e.g., $\mathcal{X}$, with the cardinality of a set being $|\mathcal{X}|$. We denote vectors and matrices with boldface lowercase and uppercase letters, e.g., vector $\vect{x}$ and matrix $\mat{X}$. The single matrix element in the $n$th row and the $k$th column of the matrix $\mat{X}$ is denoted as $x_{nk}$. The transpose and conjugate transpose of a matrix $\mat{X}$ are written as $\mat{X}^\top$ and $\mat{X}^H$, respectively, while the Hadamard product is indicated with the operator $\odot$ and the outer product with the operator $\otimes$. The diagonal matrix $\mat{D}$ with diagonal entries $\vect{d}$ is denoted as $\text{diag}(\vect{d})$ and the all-one column vector of length $N$ is denoted as $\boldsymbol{1}_N$.
A complex normal distribution with mean $\mu$ and variance $\sigma^2$ is denoted as $\mathcal{CN}(\mu,\sigma^2)$. Random variables are denoted with sans-serif font, e.g. $\mathsf{x}$, with multivariate random variables $\mvrv{x}$, mutual information $I(\mathsf{x}_1,\mathsf{x}_2)$, entropy $H(\mathsf{x})$ and cross-entropy $H(\mathsf{x}_1 ||\mathsf{x}_2)$.

\section{System Model}\label{sec:sysmodel}
In this paper, we investigate a monostatic \ac{JCAS} setup, where the transmitter and the sensing receiver are co-located, e.g., part of the same base station, and have multiple antennas. Our goal is to detect a target and estimate its \ac{AoA} $\theta$ based on the reflection of the transmitted signal at the potential target. The transmit signal is simultaneously used to communicate with a \ac{UE} with a single antenna at a different position than the sensing target. The \ac{UE} is located randomly at an azimuth angle $\varphi \in [\varphi_{\min},\varphi_{\max}]$ and the radar target is located randomly at an azimuth angle of ${\theta} \in[\theta_{\min},\theta_{\max}]$. In the considered scenario, the communication area $[\varphi_{\min},\varphi_{\max}]$ and sensing area $[\theta_{\min},\theta_{\max}]$ do not overlap. 
We consider multi-snapshot sensing with $N_{\text{win}}$ samples to provide more detailed information on the sensing targets. The system block diagram is illustrated in Fig. \ref{fig:flowgraphtrain_mono}. The blocks shaded in blue are trainable \acp{NN} and the following subsections explain in detail the signal processing of the system.

\subsection{Transmitter}
A modulator with $M =2^c$, $c \in \mathbb{N}$ different modulation symbols transforms the data symbols $m\nobreak \in \nobreak \{1,2,\ldots,M\}$ into complex symbols $x \in \mathcal{M} \subset \mathbb{C}$, where $\mathcal{M}$ contains the $M$ distinct modulation symbols. The transmitter generates a 
sequence of $c N_{\text{win}}$ independent and identically distributed random bits that are transformed into a vector $\vect{x}\in \mathbb{C}^{N_{\text{win}}}$ through a fixed bit mapping for blockwise processing. Each bit vector $\vect{b}=(b_1, \ldots, b_{c})^\top \in \{0,1\}^{c}$ is mapped to a unique symbol $x$.

The transmitter is equipped with $K$ antennas and uses digital precoding, i.e., a unique complex factor $v_k = g_k\exp(\mathrm{j}\vartheta_k)$ is generated for each antenna $k \in \{1,2,\ldots,K\}$ with amplitude $g_k$ and phase shift $\vartheta_k$ to steer the signal to a certain area of interest. The beamformer is implemented as an \ac{NN} and has the limits of the azimuth angle regions in which communication and sensing should take place, i.e., $\{\varphi_{\min},\varphi_{\max}, \theta_{\min},\theta_{\max}\}$ as inputs. The use of an \ac{NN} enables analysis of an optimized power trade-off between the sensing and communication functionality. The output of the \ac{NN} is a vector $\vect{v}\in \mathbb{C}^{K} = \left(v_1, v_2, \ldots , v_K \right)^\top$ containing the complex factors of each antenna.
The modulator and beamformer employ power normalization to meet power constraints.

The transmit signal $\mat{Y}\in \mathbb{C}^{K \times N_{\text{win}}}$ is the outer product of the complex modulation symbols $\vect{x}$ and the beamformer output $\vect{v}$. Specifically, $\mat{Y}$ can be expressed as
\begin{align}
	\mat{Y} = \vect{v} \otimes \vect{x}.\label{eq:beam}
\end{align}

\begin{figure*}[!t]
\normalsize
\begin{align}
\label{eq:CRB}
    {C}_{\text{CR}} (\theta) &= \frac{1}{\pi^2 \cos(\theta)^2}\frac{\sigma_{\text{ns}}^2}{2N_{\text{win}}} \left( \Re \left\{  \left[ \beta_{\text{s}} \sigma_{\text{s}}^2 \left[\frac{K \beta_{\text{s}} \sigma_{\text{s}}^2}{\sigma_{\text{ns}}^2 + K\beta_{\text{s}} \sigma_{\text{s}}^2} \right]\right] \cdot  \frac{K^3 - K}{12} \right\}\right)^{-1} \tag{9}
\end{align}
\hrulefill
\vspace*{-0.3cm}
\end{figure*}

\subsection{Channels}
A part of the radiated power is steered toward the communication receiver by the beamformer while another part is reflected by the object of interest and reaches the sensing receiver co-located with the transmitter.
The signal propagation from $K$ antennas towards the object located at an azimuth angle $\varphi_i$ is modeled with the spatial angle matrix $\mat{a}_{\text{TX}}(\vect{\varphi})=(\vect{a}_{\text{TX}}(\varphi_{1})\, \ldots\, \vect{a}_{\text{TX}}(\varphi_{N_{\text{win}}})) \in \mathbb{C}^{K \times N_{\text{win}}}$ whose columns are given by
\begin{align}
	 \vect{a}_{\text{TX}}(\varphi) = \left(1,\text{e}^{\mathrm{j}\pi  \sin\varphi},\ldots, \text{e}^{\mathrm{j}\pi (K-1) \sin\varphi}\right)^\top,
\end{align}
assuming the antenna spacing matches exactly $\lambda/2$ of the transmission wavelength $\lambda$.

For communications, we consider the \ac{UE} at a different uniformly distributed angle for each communication sample to generate varied training data. The signal $\mat{y}$ experiences a single-tap Rayleigh fading channel before being received by the \ac{UE} with a single antenna as
\begin{align}
	\vect{z}_{\text{c}} =  \boldsymbol{1}^\top_K \left(\mat{a}_{\text{TX}}(\vect{\varphi}) \odot \mat{y}\right) \text{diag}(\vect{\alpha}_{\text{c}}) + \vect{n_{\text{c}}}^{\top},
\end{align}
with channel coefficient $\alpha_{\text{c},n} \sim \mathcal{CN}(0,\sigma_\text{c}^2)$ and noise samples $n_{\text{c},n}\nobreak \sim \nobreak \mathcal{CN}(0,\sigma_\text{nc}^2)$. 
We select a constant $\varphi_i \in [\varphi_{\min},\varphi_{\max}]$ uniformly at random during an observation window $N_{\text{win}}$ to analyze multi-snapshot sensing and define $\text{SNR}_{\text{c}}= \sigma_\text{c}^2 /{\sigma_\text{nc}^2}$. The total \ac{SNR} is corrected with the beamforming gain $\beta_{\text{c}}$ to $\text{SNR}=\nobreak \beta_{\text{c}} \cdot \text{SNR}_{\text{c}}$.

With $T\in \nobreak \{0,1\}$ indicating the presence of a potential target, we express the sensing signal reflected from said target in the monostatic setup as
\begin{align}
	\mat{z}_{\text{s}} = T \vect{a}_{\text{RX}}({\theta}) \vect{a}_{\text{TX}}({\theta})^\top \mat{y} \text{diag}(\vect{\alpha}_{\text{s}}) + \mat{N_{\text{s}}},
\end{align}
with the radar target following a Swerling-1 model~\cite{Swerling1960} with ${\alpha}_{\text{s},n}\nobreak \sim \mathcal{CN}(0,\sigma_{\text{s}}^2)$ representing the radar cross section and path loss and $\mat{N}_{\text{s}}$ consisting of noise samples $n_{\text{s},nk} \sim \mathcal{CN}(0,\sigma_{\text{ns}}^2)$. The spatial angle vectors are $\vect{a}_{\text{RX}}({\theta})=\vect{a}_{\text{TX}}({\theta})$, with $\theta$ being the \ac{AoA} of the target throughout the observation window $N_{\text{win}}$. With a Swerling-1 model, we model scan-to-scan deviations of the \ac{RCS}, which manifest as a change in $\alpha_{\text{s},n}$ during $N_{\text{win}}$. The radial velocity of the target is assumed to be zero, so no Doppler shift occurs.

\subsection{Sensing Receiver}
The sensing receiver, which is part of the base station, detects the potential target and estimates its \ac{AoA} using a linear array of $K$ antennas. 
We consider multiple snapshot sensing with $N_{\text{win}}$ snapshots, enabled by forming the short-term spatial auto-correlation of all samples considered across the receive antennas defined as
\begin{align}
    \text{Corr}(\mat{z}_{\text{s}},\mat{z}_{\text{s}}) := \frac{1}{N_{\text{win}}}  \mat{z}_{\text{s}} \mat{z}_{\text{s}}^H \quad \in \mathbb{C}^{K\times K}.
\end{align}
This metric approximates a sufficient statistic for \ac{AoA} estimation and detection, as outlined in Appendix~\ref{app:suffstat}.
Next, $\text{Corr}(\mat{z}_{\text{s}},\mat{z}_{\text{s}})$ and $N_{\text{win}}$, and $\sigma_{\text{n,s}}$ are passed to the target detection and angle estimation blocks which are implemented as \acp{NN} and described in Sec.~\ref{sec:NN}. The target detection block outputs a probability $p_{\text{T}} \in [0,1]$ that denotes the certainty that a target is present ($T=1$).
The angle estimation block outputs $\hat{{\theta}} \in [- \frac{\pi}{2}, \frac{\pi}{2}]$, which indicates the estimated azimuth \ac{AoA} of the target.

\subsection{Communication Receiver}
At the communication receiver, our goal is to recover the transmitted bits based on the received signal, i.e., demodulate the received signal. The communication receiver is implemented as a \ac{NN} that outputs \acp{LLR} $\ell \in \mathbb{R}^{n \times N_{\text{win}}}$ that can be used as input to a soft decision channel decoder. For the demodulation, we assume that channel estimation has already been performed at the communication receiver and that the precoding matrix $\vect{v}$ is known. Therefore, \ac{CSI} $\vect{\gamma}=\nobreak \vect{v}^{\top} \mat{a}_{\text{TX}}(\vect{\varphi})\text{diag}(\vect{\alpha}_{\text{c}})$ is available at the communication receiver. It is important to note that this \ac{CSI} has no effect on the sensing functionality of the system.

\subsection{Performance Indicators}
We formulate bounds on the communication throughput and estimation accuracy as theoretical performance indicators for the system.
\subsubsection{Constellation Kurtosis}\label{sec:kurtosis}
In~\cite{Geiger25}, it was shown that the random-deterministic sensing trade-off for OFDM-sensing can be controlled by the kurtosis, i.e., the fourth standardized moment of the constellation, if a matched filter is applied to estimate the sensing channel. As OFDM sensing and \ac{AoA} estimation are both harmonic retrieval problems, these findings can be reapplied to our scenario. The constellation kurtosis is directly connected to the achievable detection rate in sensing. 
The kurtosis is not directly linked to the communication performance. Nevertheless, the kurtosis constraint places restrictions on constellation design and can affect achievable data rates. Assuming equal symbol probability, we use the mean minimum distance $\bar{d}_{\min}$ defined as
\begin{align}
    \bar{d}_{\min} = \frac{1}{|\mathcal{M}|}\sum_{x_m \in \mathcal{M}} \min_{x_n \in \mathcal{M} \backslash \{x_m\}} |x_n-x_m|
\end{align}
as an indicator for the communication performance and relate it to the kurtosis of a constellation. Typically, a larger $\bar{d}_{\min}$ leads to a smaller \ac{BER}, especially for pragmatic constellation diagrams where the spacing between neighboring constellation points is similar for all constellation symbols.  

\begin{figure}
    \centering
    \begin{tikzpicture}[>=latex]
    
    \draw[dashed, KITblue] (0,0) circle [radius=0.8];
    \draw[dashed, KITpurple] (0,0) circle [radius=1.2];

    \draw[gray] (-45:0.8) circle [radius=0.275];
    \draw[gray] (-45-22.5:1.2) circle [radius=0.275];
    \draw[gray] (-90:0.8) circle [radius=0.275];
    \draw[gray] (-90-22.5:1.2) circle [radius=0.275];

    \draw[gray] (45:0.8) circle [radius=0.275];
    \draw[gray] (45-22.5:1.2) circle [radius=0.275];
    \draw[gray] (90:0.8) circle [radius=0.275];
    \draw[gray] (90-22.5:1.2) circle [radius=0.275];

    \draw[gray] (135:0.8) circle [radius=0.275];
    \draw[gray] (135-22.5:1.2) circle [radius=0.275];
    \draw[gray] (180:0.8) circle [radius=0.275];
    \draw[gray] (180-22.5:1.2) circle [radius=0.275];

    \draw[gray] (225:0.8) circle [radius=0.275];
    \draw[gray] (225-22.5:1.2) circle [radius=0.275];
    \draw[gray] (0:0.8) circle [radius=0.275];
    \draw[gray] (0-22.5:1.2) circle [radius=0.275];

    \fill[KITgreen] (-45:0.8) circle (2pt);
    \fill[KITgreen] (-45-22.5:1.2) circle (2pt);
    \fill[KITgreen] (-90:0.8) circle (2pt);
    \fill[KITgreen] (-90-22.5:1.2) circle (2pt);

    \fill[KITgreen] (45:0.8) circle (2pt);
    \fill[KITgreen] (45-22.5:1.2) circle (2pt);
    \fill[KITgreen] (90:0.8) circle (2pt);
    \fill[KITgreen] (90-22.5:1.2) circle (2pt);

    \fill[KITgreen] (135:0.8) circle (2pt);
    \fill[KITgreen] (135-22.5:1.2) circle (2pt);
    \fill[KITgreen] (180:0.8) circle (2pt);
    \fill[KITgreen] (180-22.5:1.2) circle (2pt);

    \fill[KITgreen] (225:0.8) circle (2pt);
    \fill[KITgreen] (225-22.5:1.2) circle (2pt);
    \fill[KITgreen] (0:0.8) circle (2pt);
    \fill[KITgreen] (0-22.5:1.2) circle (2pt);

    \draw[KITpurple] (0,0) -- (225-22.5:1.2);
    \draw[KITblue] (0,0) -- (225:0.8);
    \draw[fill=KITred!40] (0,0) -- (225-22.5:0.3) arc [start angle=225-22.5, delta angle=22.5, radius=0.3] -- node[near end, below,xshift=2,yshift=1] {\tiny{$\frac{2\pi}{M}$}} (0,0);

    \draw[red!80!black] (225:0.8) --  (225-22.5:1.2) ;
    \draw[red!80!black] (225-22.5:1.2) -- (180:0.8) -- (180-22.5:1.2) -- (135:0.8);

    \node[red!80!black] at (-180:1.5) {\small $\bar{d}_{\min}$};
    \node[KITpurple] at (0:1.5) {$R_1$};
    \node[KITblue] at (45:0.4) {$R_2$};

    \end{tikzpicture}
    \caption{Example constellation of A\ac{PSK} to explore performance for constellations with different kurtosis}
    \label{fig:apsk}
\end{figure}
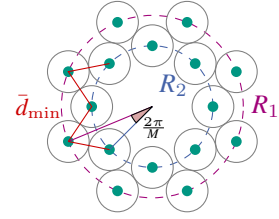

To demonstrate the relationship between the minimum distance $\bar{d}_{\min}$ and the kurtosis $\kappa$, we design an \ac{APSK} as a constellation using two 8 \acp{PSK} on radii $R_1$ and $R_2$ with an offset phase of $\frac{2\pi}{M}$ between the constellation points on circle $R_1$ and $R_2$ as shown in Fig.~\ref{fig:apsk}.
The \ac{APSK} allows a continuous evaluation of the kurtosis. 

Then, $R_1=\sqrt{2-R_2^2}$ is obtained by power normalization and the kurtosis $\kappa$ of the constellation with mean $\mu_{\mathcal{M}}=0$ and standard deviation $\sigma_{\mathcal{M}}=1$ follows as
\begin{align}
    \kappa = \frac{1}{|\mathcal{M}|}\sum_{m \in \mathcal{M}} \left| \frac{x_m - \mu_{\mathcal{M}}}{\sigma_{\mathcal{M}}} \right|^4  = \frac{R_1^4+R_2^4}{2}.
\end{align}

The minimum distance  follows from trigonometric identities as
\begin{align}
    \bar{d}_{\min} = \sqrt{2-2\sqrt{\kappa}\cos\left(\frac{2\pi}{M}\right)},
\end{align}
which is valid for $\bar{d}_{\min}\leq 2R_2\sin\left(\frac{2\pi}{M}\right)$, i.e., until the inner circle becomes too small to fit an 8\ac{PSK} with point distance of at least $\bar{d}_{\min}$.
This constellation yields an achievable relationship between kurtosis and mean minimum distance for this specific constellation topology without claiming optimality.

\subsubsection{Cramér-Rao Bound} 
The adaptation of the \ac{CRB} to the designed system is derived in Appendix~\ref{app:crb}.
For the estimation of the angle of a single target, the \ac{CRB} amounts to \eqref{eq:CRB}, where $N_{\text{win}}$ is the number of samples collected, $\theta$ the angle to be estimated, $\beta$ the beamforming gain, and $K$ the number of antennas~\cite{Trees2002}. The given \ac{CRB} is a lower bound for the variance of an unbiased estimator.

\subsubsection{Bit-wise Mutual Information}
Although a blockwise processing is indicated in Fig.~\ref{fig:flowgraphtrain_mono} with block length $N_{\text{win}}$, we consider the statistics of single modulation symbols for evaluation (as the channel is memoryless). 
The aim of the communication receiver is to maximize the \ac{BMI}, since transmission uses \ac{BICM} and \ac{FEC} to enable reliable communication. The \ac{BMI} is therefore used as a surrogate measure for communication performance after \ac{FEC} without the need to consider a specific coding approach in this work. Maximizing the \ac{BMI} is equivalent to minimizing the \ac{BCE} between true bit labels $\mathsf{b}_i$ and the estimated bit labels $\mathsf{\hat{b}}_i$~\cite[Eq.~(8)]{Cammerer2020}:
\begin{align}
\setcounter{equation}{9}
    \text{BMI}%
    &= \sum_{i=1}^{\log_2 M} I(\mathsf{b}_i;\mathsf{z}_{\text{c}})\\
    &= H(\mvrv{B}) - \sum_{i=1}^{\log_2M}H(\mathsf{b}_i||\mathsf{\hat{b}}_i) \notag \\ &\qquad+ \sum_{i=1}^{\log_2 M} 
    \mathbb{E}_{\gamma}\left[ D_{\text{KL}}(p(\mathsf{b}_i|\mathsf{z}_{\text{c}}) || {p}(\mathsf{\hat{b}}_i|\mathsf{z}_{\text{c}}) )\right],
\end{align}
with source entropy $H(\mvrv{B})$, binary cross-entropy $H(\mathsf{b}_i||\hat{\mathsf{b}}_i)$ and the expected \acl{KLD} $D_{\text{KL}}$ that denotes the expected mismatch of the true posterior $p(\mathsf{b}_i|\mathsf{z}_{\text{c}})$ given the observed receiver signal and the approximated (by the receiver) posterior $p(\mathsf{\hat{b}}_i|\mathsf{z}_{\text{c}})$ over different channel realizations. 
The \ac{BMI} is an achievable rate using binary coding and pragmatic coded modulation.

\section{Neural Network Training and Validation}\label{sec:NN}

\subsection{Neural Network Configuration}\label{sec:NNconfig}
Our system is configured similarly to that of \cite{Muth2023}. Function blocks such as beamforming, target detection, angle estimation, demapping and optionally modulation are implemented as separate \acp{NN}. The \ac{NN} layer dimensions are given in Tab.~\ref{tab:NNsizes}. 
The \acp{NN} consist of fully connected layers with ELU activation function in the hidden layers.

We implement modulation as a classical \ac{QAM} and for comparison as a trainable \ac{NN}. Inputs are symbol indices with a known fixed bit mapping.
The inputs of the beamformer are the areas of interest for sensing $\{\theta_{\min}, \theta_{\max}\}$ and for communication $\{\varphi_{\min}, \varphi_{\max}\}$. The output of the transmitter is subject to power normalization.

At the communication receiver, \ac{MMSE} equalization is performed, compensating for the complex random channel tap, to achieve better convergence along different \ac{SNR} values. The outputs of the communication receiver are interpreted as \acp{LLR} for each bit. For the \ac{BER} calculation, we use the hard decision of these \ac{LLR} values.
As stated in Sec.~\ref{sec:sysmodel}, we strive to improve the performance for multiple snapshot estimation by calculating the short-term spatial \ac{ACM} $\text{Corr}(\mat{Z}_{\text{s}},\mat{Z}_{\text{s}})$.
The number of input neurons of the sensing receiver is $K^2+2$.
Two input neurons have $N_{\text{win}}$ and $\sigma_{\text{ns}}$ as inputs and allow the investigation of sensing for varying channel parameters. Specifically, our systems are trained for generalized $N_{\text{win}}$ and $\sigma_{\text{ns}}$, allowing flexible investigation within the range of training parameters. This parameterization leads to roughly the same communication and sensing performance as systems trained individually for different $N_{\text{win}}$ and $\sigma_{\text{ns}}$, while allowing flexible operation without requiring a change of the \ac{NN} weights and, at the same time reducing the computational complexity required for training.

During training, a vector indicating the presence of a target is fed to the sensing receiver in order to calculate the detection threshold that is needed to keep the false alarm rate $P_{\text{f}}$ constant. This threshold is added to the output of the detection \ac{NN} before applying the output sigmoid function. Since this threshold is numerically calculated for each system, there are small variations expected. The output function of the angle estimation \ac{NN} is $\frac{\pi}{2}\tanh(\cdot)$, normalizing the output to $\pm \frac{\pi}{2}$.

\begin{table}[t]
\centering
\caption{Sizes of neural networks}
\begin{tabular}{@{}lccc@{}}
\toprule
Component  & Input layer & Hidden layers &  Output\\ 
\midrule
Beamformer   & $4$      & $\{ K,K,2K \}$ & $2K$  \\
Demapper   & $3$      & $\{10M,10M,10M,10M\}$          & $\log_2(M)$  \\
Angle estimation   &   $2K^2+2$    & $\{8K,4K,4K,K\}$          & $1$  \\
Detection   & $2K^2+2$       & $\{2K,2K,K\}$          & $1$  \\
\midrule
Modulator   & $M$      & $\{ 8M,8M,8M \}$ & $2$  \\
\bottomrule
\end{tabular}
\label{tab:NNsizes}
\vspace{-0.3cm}
\end{table}

\subsection{Loss Functions}
There are three main components of the loss function, resulting in a multi-objective optimization, which evaluates the performance of communication, detection, and angle estimation.
We introduce a weight $w_{\text{s}} \in [0,1]$ that controls the impact or perceived importance of the sensing tasks, resulting in the full loss term
\begin{align}
    L = (1-w_{\text{s}})L_{\text{comm}} + w_{\text{s}}L_{\text{detect}} + w_{\text{s}}L_{\text{angle}} \label{eq:loss}.
\end{align}
The weight $w_{\text{s}}$ affects both the power allocation and if trainable, also the constellation shape.
\ac{JCAS} systems have been trained in~\cite{Muth2023} with $N$ different scenarios using the loss given by 
\begin{multline}
    L =  (1-w_{\text{s}})\underbrace{\sum_{i=1}^{\log_2M}H(\mathsf{b}_i||\hat{\mathsf{b}}_i)}_{L_{\text{comm}}} \\+ w_{\text{s}} \left( \underbrace{H(\mathsf{t}||\hat{\mathsf{t}})}_{L_{\text{detect}}} + \underbrace{\frac{1}{N}\sum_{i=1}^{N} \mathbf{I}_{t_i=1}(\theta_i-\hat{\theta_i})^2}_{L_{\text{angle}}} \right) \label{eq:lossalt},
\end{multline}
with indicator function $\mathbf{I}_{(\cdot)}$.
When training multiple functionalities and multiple operating scenarios simultaneously, we observed a reduced performance when using~\eqref{eq:lossalt} as a loss function. Especially the \ac{AoA} estimation showed unreliable convergence. The achievable precision, which is bounded by~\eqref{eq:CRB}, depends significantly on the chosen $N_{\text{win}}$ and $\sigma_{\text{ns}}$.  After training, the estimator should yield a \ac{MSE} similar to the lower bounds, i.e., the \ac{CRB}, of the estimator. The lower bound is not constant in all training scenarios, since it depends on $N_{\text{win}}$ and $\sigma_{\text{nc}}$ resulting in a perturbation of the loss. Therefore, we introduce bound-informed adaptations to ensure a robust performance over a range of \acp{SNR} and observation window lengths $N_{\text{win}}$.
We expect increased precision and better convergence behavior if the loss is normalized across multiple \acp{SNR} and $N_{\text{win}}$.

The \ac{CRB} is used for an informed modification of the loss function used for the training of \acp{NN}. Under the assumption of $\sigma_{\text{ns}}^2 \ll K\beta \sigma_{\text{s}}^2$, we simplify~\eqref{eq:CRB} to separate the expression into parts depending on $N_{\text{win}}$ and $\sigma_{\text{ns}}$:
\begin{align}
    C_{\text{CR}} (\theta) &\approx \frac{1}{\pi^2 \cos(\theta)^2} \frac{\sigma_{\text{ns}}^2}{N_{\text{win}}}   \frac{1}{\beta_{\text{s}} \sigma_{\text{s}}^2}   \frac{6}{K^3 - K}.
\end{align}
The factor ${\sigma_{\text{ns}}^2}/{N_{\text{win}}}$ describes the impact of the observation window and \ac{SNR} on the bound. Therefore, we modify the loss term with the correction factor ${N_{\text{win}}}/{ \sigma_{\text{ns}}^2}$. The proposed loss term is then given by
\begin{align}
    L_{\text{angle}} =& \frac{1}{N} \sum_{i=1}^{N} \mathbf{I}_{t_i=1} \frac{N_{\text{win},i}}{ \sigma_{\text{ns},i}^2}(\theta_i-\hat{\theta}_i)^2,\label{eq:anglelos}
\end{align}
achieving loss terms with similar magnitude for varying $N_{\text{win}}$ and $\sigma_{\text{ns}}$ and regularizing the output.

There are multiple examples for training communication systems over varying \acp{SNR}~\cite{Rode22,Cammerer2020}; but a gradient perturbation or problems during training have not been reported. To complete our analysis of loss behavior, we consider the behavior of \ac{BCE} for our channel model and varying $N_{\text{win}}$ and \acp{SNR}. In the following, we show why a modification of the communication loss term is not necessary.
We adapt the communication loss that is evaluated based on the known bits $\mathsf{b}_{i,n}$, with $i$ denoting the bit in a specific symbol and $n$ denoting the symbol index, and estimated bits $\mathsf{\hat{b}}_{i,n}$ 
 based on output probabilities $p(\mathsf{\hat{b}}_{i,n}=1|z_{\text{c}})=\sigma(\ell_{i,n})$, with $\sigma(\cdot)$ being the sigmoid function and the output \acp{LLR} $\ell_{i,n}$. Traditionally, the \ac{BCE} $ H(\mvrv{b}|| \hat{\mvrv{b}})$ is used as a loss function, given by
\begin{align}
    \sum_{i=1}^{\log_2M}H(\mathsf{b}_i||\hat{\mathsf{b}}_i) =&  - \sum_{i=1}^{\log_2 M} \mathsf{b}_i \log_2(p(\mathsf{\hat{b}}_i=1|z_{\text{c}}))\notag\\
    &+ (1-\mathsf{b}_i) \log_2(1-p(\mathsf{\hat{b}}_i=1|z_{\text{c}})).
\end{align}
Assuming independent bits, the \ac{BCE} and \ac{BMI} are related as~\cite{Cammerer2020}
\begin{align}
    \sum_{i=1}^{\log_2M}H(\mathsf{b}_i||\hat{\mathsf{b}}_i) =& H(\mvrv{B}) - \text{BMI}%
    \notag \\ &+ \sum_{i=1}^{\log_2 M} \mathbb{E}_{\gamma}\left[ D_{\text{KL}}(p(\mathsf{b}_i|z_{\text{c}}) || {p}(\mathsf{\hat{b}}_i|z_{\text{c}}) )\right].
\end{align}
We lower bound the \ac{BCE} as 
\begin{align}
\sum_{i=1}^{\log_2M}H(\mathsf{b}_i||\hat{\mathsf{b}}_i) \stackrel{(a)}{\geq}& H(\mvrv{B}) -I(\mvrv{b};{{z_{\text{c}}}}) \notag\\ 
    &+ \sum_{i=1}^{\log_2 M} \mathbb{E}_{\gamma}\left[ D_{\text{KL}}(p(\mathsf{b}_i|z_{\text{c}}) || {p}(\mathsf{\hat{b}}_i|z_{\text{c}}) )\right] \label{eq:kl-term}\\
    \stackrel{(b)}{\geq}& H(\mvrv{B}) -I(\mvrv{b};{{z_{\text{c}}}}) \\
    \stackrel{(c)}{\geq}& H(\mvrv{B}) - C,
\end{align}
where in $(a)$, we upper bound the \ac{BMI} by the mutual information $I(\mvrv{b};{{z_{\text{c}}}})$.
The \ac{KLD} $D_{\text{KL}}\geq 0$ describes the mismatch of the true posterior distributions and the distributions at the \ac{NN} output. With converged systems, this mismatch should be very small, encouraging us to drop the term in $(b)$.
Finally, we bound the mutual information by the channel capacity $C$ in $(c)$.

The ergodic channel capacity $C$ for a Rayleigh fading channel is given by $C_{\text{e}}=\log_2\left(1+\frac{\sigma_{\text{c}}^2 \cdot \bar{\beta}_\text{c} P_s}{\sigma_{\text{nc}}^2}\right)$~\cite{Li2005}, with $\bar{\beta}_\text{c}$ denoting the average beamforming gain for the area of interest for communication. For a single realization, $I(\mvrv{B};z_{\text{c}})$ could be larger than the ergodic capacity, but considering many different realizations for a loss function and the law of large numbers diminish this possibility.
Then
\begin{align}
     \sum_{i=1}^{\log_2M}H(\mathsf{b}_i||\hat{\mathsf{b}}_i) \gtrapprox& H(\mvrv{B})  - \log_2\left(1+\frac{\sigma_{\text{c}}^2 \bar{\beta}_\text{c} P_s}{\sigma_{\text{nc}}^2}\right)
     \\
     =& H(\mvrv{B})  - \log_2\left(\underbrace{\sigma_{\text{nc}}^2+\sigma_{\text{c}}^2 \bar{\beta}_\text{c} P_s}_{\sigma_{\text{nc}}^2 \ll \sigma_{\text{c}}^2 \bar{\beta}_\text{c} P_s}\right) \notag \\
     &\qquad+\log_2\left(\sigma_{\text{nc}}^2\right)  \\
     \approx& H(\mvrv{B}) - \log_2\left({\sigma_{\text{c}}^2 \bar{\beta}_\text{c} P_s}\right) +\log_2\left(\sigma_{\text{nc}}^2\right).\label{eq:commapprox}
\end{align}
The approximation in~\eqref{eq:commapprox} is valid for high \ac{SNR}. In lower \ac{SNR} scenarios,  the \ac{BCE} is underestimated when using~\eqref{eq:commapprox}. Based on~\eqref{eq:commapprox}, the terms $\log_2\left({\sigma_{\text{c}}^2 \bar{\beta}_\text{c} P_s}\right)$ describes the impact of the beamforming while the term $\log_2\left(\sigma_{\text{nc}}^2\right)$ represents the impact of \ac{SNR} variations which we aim to remove from the loss. A loss function modification following our arguments for~\eqref{eq:anglelos} results in:
\begin{align}
     L_{\text{comm}} &=  \frac{-1}{N N_{\text{win}} \log_2 M}  \sum_{i=1}^{N \cdot N_{\text{win}}} \left( \sum_{k=1}^{\log_2 M} \right.\notag \\
     &\left( \mathsf{b}_{i,k} \cdot \log_2({p}(\mathsf{\hat b}_{i,k}=1|z_{\text{c},i}))+(1- \mathsf{b}_{i,k}) \right.\notag\\
    &\left.  \cdot \log_2({p}(\mathsf{\hat b}_{i,k}=0|z_{\text{c},i})) \vphantom{\sum_{i=1}^N} \Bigr) +\log_2(\sigma_{\text{nc},i}^2)\right). \label{eq:commloss}
\end{align}
This loss function can be used for optimization of the various \ac{NN}-based blocks using the Adam optimizer~\cite{Kingma2014}; therefore, the gradients of $L_{\text{comm}}$ with respect to the weights of the neural networks are used for optimization. However, since the offset term $\log_2(\sigma_{\text{nc},i}^2)$ in~\eqref{eq:commloss} is additive and does not depend on the weights, its gradient will be zero and will not affect the training. Hence, it can be omitted from~\eqref{eq:commloss}, which then boils down to the \ac{BCE} loss, showing that modification of the \ac{BCE} loss for communication is not necessary.

For the detection loss, we can use the same approach of modifying the \ac{BCE} between the true presence of targets $\mathsf{t}~\in~\{0,1\}^N$ and the detection output $p_{\text{T}}$, keeping in mind that the \ac{SNR} scales with $\sqrt{N_{\text{win}}}$:
\begin{align}
    H(\mathsf{t}||\hat{\mathsf{t}}) & \approx H(\mathsf{t})  - \log_2\left({\sigma_{\text{s}}^2 \bar{\beta}_\text{s} P_s}\right) +\log_2\left(\frac{\sigma_{\text{ns}}^2}{\sqrt{N_{\text{win}}}}\right).
\end{align}
We can derive a loss similar to~\eqref{eq:commloss}, given by
\begin{align}
    L_{\text{detect}} =& \frac{-1}{N} \sum_{i=1}^{N} \mathsf{t}_{i} \cdot \log_2(p_{\text{T}})
    +(1-\mathsf{t}_{i}) \cdot \log_2(1-{p_{\text{T}}})\notag\\
    &+ \log_2\left(\frac{\sigma_{\text{ns,}i}^2}{\sqrt{N_{\text{win},i}}}\right)\label{eq:detectloss}.
\end{align} 
With the same arguments as given before, the last term $\log_2\left(\frac{\sigma_{\text{ns,}i}^2}{\sqrt{N_{\text{win},i}}}\right)$ does not impact the training behavior. Therefore, a modification of the \ac{BCE} as loss term for communication and object detection is not necessary for the described system configuration.

Furthermore, we want to ensure a constant false alarm rate $P_{\text{f}}$ over different \acp{SNR} and observation window lengths $N_{\text{win}}$. The input of the detection \ac{NN} is normalized to $\text{Corr}(\mat{z}_{\text{s}},\mat{z}_{\text{s}})/\sigma_{\text{ns},i}^2$ to obtain the same noise statistic and, therefore, false alarm rates for different \acp{SNR}. We apply different decision thresholds for each window length $N_{\text{win}}$ that are calculated numerically after training the \ac{NN} components, explained in Sec.~\ref{sec:nn-train}.

\subsection{Neural Network Training}\label{sec:nn-train}
The training comprises three phases: Pre-training, fine-tuning, and limiting. In pre-training, the detection and angle estimation neural network are trained independently, i.e., the other loss terms of Eq.~\ref{eq:loss} are set to $0$. We use a total of $2.5\cdot\nobreak10^7$ communication symbols for both pre-training steps, divided into mini-batches of $10^4$ symbols. We use the Adam optimizer with a learning rate of $10^{-4}$. The length of the sensing window is randomly and uniformly chosen between $1$ and $15$ for each sensing state, to generalize to different $N_{\text{win}}$ and to give insight into multi-snapshot behavior. Fine-tuning establishes the operating point of the \ac{JCAS} trade-off. The fine-tuning is performed on $5\cdot 10^7$ symbols by using the whole loss function in ~\eqref{eq:loss} starting with the parameters established in the pre-training. We use the same hyperparameters as used for pre-training. Lastly, limiting ensures that the constant false alarm rate is maintained. In the limiting phase, the system runs separately for $10^4$ symbols for each length of the sensing window $N_{\text{win}}$. The neural network components are not trained anymore in this phase, but the decision threshold for detection is refined as described in~\cite{Muth2023}.

For validation of the communication component, we choose the \ac{BMI} and \ac{BER} as metrics. The \ac{BMI} represents the maximum number of bits per symbol that can be reliably transmitted on average with the given system.

The object detection task is evaluated on the basis of its detection rate and false alarm rate. For direct comparison, we design detectors with a constant false alarm rate.

The \ac{AoA} estimation is evaluated on the \ac{RMSE} of angle estimates. We evaluate the \ac{AoA} on all the present targets instead of only on the detected targets as in \cite{MateosRamos2021}.

\section{Simulation Results and Discussion}\label{sec:sims}
In our simulations\footnote{The source code for simulations performed for this paper can be found at \url{https://github.com/frozenhairdryer/JCAS-loss-shape-precode}.}, the communication receiver is located at an \ac{AoA} of $\varphi \in [30^\circ,50^\circ]$. Radar targets are found in a range $\theta \in [-20^\circ,20^\circ]$.
Our monostatic transmitter and sensing receiver are both simulated as a linear array with 16 antennas and we consider an observation window up to $N_{\text{win}}=15$ and modulation with $M=16$. For the radar receiver, our objective is to achieve a false alarm rate of $P_{\text{f}}=\nobreak10^{-2}$ while optimizing the detection rate and angle estimator.

\subsection{Benchmarks for Modulation, Detection and Angle Estimation}
\acused{ESPRIT}
We train \acp{NN} to optimize the modulation format as in~\cite{Muth2023}, enabling geometric constellation shaping, and compare it to a classical \ac{QAM} to quantify the gains of the optimized constellation.

For detection, we employ a generalized power detector based on a \ac{NP} criterion~\cite[Chap. 10]{Trees2002} as a benchmark, distinguishing between two normal distributions of mean $0$ with different variances. The exact transmit sequence $\mat{X}$ as well as prior knowledge about the sensing area are not available. In the reference detector, the average power of all the input samples $z_{\text{s},il}$ considered for sensing is computed. The detector can be formulated as
\begin{align}
    \frac{2}{\sigma_{\text{ns}}^2}\sum_{l=1}^{N_{\text{win}}} \sum_{i=1}^{K} |z_{\text{s},il}|^2 \quad\mathop{\gtreqless}_{\hat{t}=0}^{\hat{t}=1}\quad \chi^2_{2KN_{\text{win}}}(1-P_{\text{f}}),
\end{align}
where $\chi^2_{2KN_{\text{win}}}(\cdot)$ denotes the density of the chi-squared distribution with $2KN_{\text{win}}$ degrees of freedom.
The correction factor is caused by the transformation of the problem from complex to real numbers, therefore artificially doubling the number of samples but reducing the noise variance by a factor of $\sqrt{2}$.
The benchmark detector has a constant false alarm rate along varying values of $\text{SNR}_{\text{s}}$ and $N_{\text{win}}$. 

We use the well-studied \ac{ESPRIT} algorithm as a benchmark for angle estimation as presented in~\cite{Trees2002}. \ac{ESPRIT} performs close to the \ac{CRB} for high \acp{SNR} or large observation window length $N_{\text{win}}$.

\begin{figure}
\begin{subfigure}[c]{\columnwidth}
\hspace{0.5mm}
	\begin{tikzpicture}
		\begin{axis}[
			xlabel=$w_{\text{s}}$, ylabel=$P/P_{\text{total}}$,
			grid=major,
			legend cell align={left},
			legend pos=south east,
			xmin=0,xmax=1,
			ymin=0,ymax=1,
			axis line style=thick,
                width=0.95\columnwidth,
                height=0.6\columnwidth,
                legend style={name=leg},
			]
            \addlegendimage{thick, solid, KITblue} \addlegendentry{Sens.}
		\addlegendimage{thick, solid, KITpurple} \addlegendentry{Comm.}
            \addlegendimage{thick, solid, cb-3} \addlegendentry{Sum}
            \addplot [mark=o, color=KITblue, thick, mark repeat=2] table [x = ws, y expr = {\thisrow{sens}/\thisrow{total}}]
		{figures/level/NN_loss/ws_beamgain.txt};
            \addplot [mark=o, color=KITpurple, thick, mark repeat=2] table [x = ws, y expr = {\thisrow{comm}/\thisrow{total}}]
		{figures/level/NN_loss/ws_beamgain.txt};
            \addplot [mark=o, color=cb-3, thick, mark repeat=2] table [x = ws, y expr = {(\thisrow{sens}+\thisrow{comm})/\thisrow{total}}]
		{figures/level/NN_loss/ws_beamgain.txt};
            \addplot [mark=x, color=KITblue!50, thick,  mark repeat=2, mark phase=2] table [x = ws, y expr = {\thisrow{sens}/\thisrow{total}}]
		{figures/level/QAM_loss/ws_beamgain.txt};
            \addplot [mark=x, color=KITpurple!50, thick,  mark repeat=2, mark phase=2] table [x = ws, y expr = {\thisrow{comm}/\thisrow{total}}]
		{figures/level/QAM_loss/ws_beamgain.txt};
            \addplot [mark=x, color=cb-3!50, thick,  mark repeat=2, mark phase=2] table [x = ws, y expr = {(\thisrow{sens}+\thisrow{comm})/\thisrow{total}}]
		{figures/level/QAM_loss/ws_beamgain.txt};
            \addplot[label=l1, draw= none, mark=o, thick] coordinates {(1,1.5) };
            \label{S}
             \addplot[label=dashed, draw= none, mark=x,thick,black!50] coordinates {(1,2)};
            \label{D}
\end{axis}
\node [draw,fill=white, left=1mm, anchor=south east] at (leg.south west) {\shortstack[l]{
            \ref{S} NN mod\\
            \ref{D} QAM}};
\end{tikzpicture}
\subcaption{Power distribution for areas of interest}
\label{fig:ws_comp}
\end{subfigure}\\
\begin{subfigure}[c]{\columnwidth}
\vspace{4mm}
	\begin{tikzpicture}
		\begin{semilogyaxis}[
			xlabel=Angle (deg), ylabel=$P$,
			legend entries={Sensing,Communication, Sum}, 
			legend cell align={left},
			legend pos=north west,
			xmin=-90,xmax=90,
			ymin=0.001,ymax=10,
			axis line style=thick,
                width=0.95\columnwidth,
                height=0.6\columnwidth,
                extra x ticks={-25, 25, 75,-75},
                tick align=inside,
                grid=major,
			]
            \fill[fill=KITblue, fill opacity=0.1] (axis cs: -20,0.001) rectangle (axis cs: 20,10);
            \fill[fill=KITpurple, fill opacity=0.1] (axis cs: 30,0.001) rectangle (axis cs: 50,10);
            \addlegendimage{thick, solid, cb-2} \addlegendentry{$w_{\text{s}}=0.9$}
		\addlegendimage{thick, solid, cb-3} \addlegendentry{$w_{\text{s}}=0.7$}
            \addlegendimage{thick, solid, cb-6} \addlegendentry{$w_{\text{s}}=0.4$}
            \addlegendimage{thick, solid, cb-1} \addlegendentry{$w_{\text{s}}=0.1$}
            \addplot [color=cb-2, thick] table [x expr= {\thisrow{angle}*180/pi}, y expr = {\thisrow{NN0.9}}]
		{figures/level/QAM_loss/ws_beams.txt};
            \addplot [color=cb-3, thick] table [x expr= {\thisrow{angle}*180/pi}, y expr = {\thisrow{NN0.7}}]
		{figures/level/QAM_loss/ws_beams.txt};
            \addplot [color=cb-6, thick] table [x expr= {\thisrow{angle}*180/pi}, y expr = {\thisrow{NN0.4}}]
		{figures/level/QAM_loss/ws_beams.txt};
            \addplot [color=cb-1, thick] table [x expr= {\thisrow{angle}*180/pi}, y expr = {\thisrow{NN0.1}}]
		{figures/level/QAM_loss/ws_beams.txt};
            \draw [black] (axis cs:0,0.002) node {\small Sensing};
            \draw [black] (axis cs:40,0.002) node {\small Comm.};
\end{semilogyaxis}
\end{tikzpicture}
\vspace{-2mm}
\subcaption{Beam pattern outputs for \ac{QAM}, areas of interests are shaded}
\label{fig:ws_comp_fine}
\end{subfigure}
\caption{Beamforming results for different operating points $w_{\text{s}}$ }
\vspace{-0.3cm}
\end{figure}
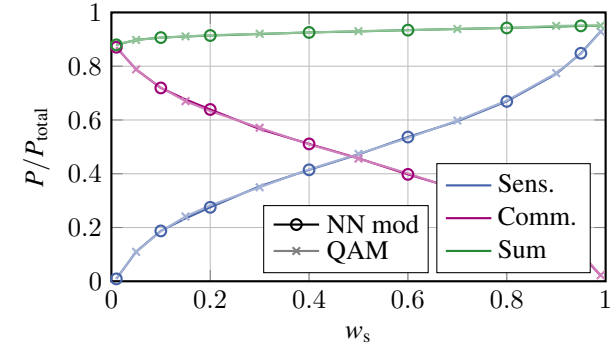
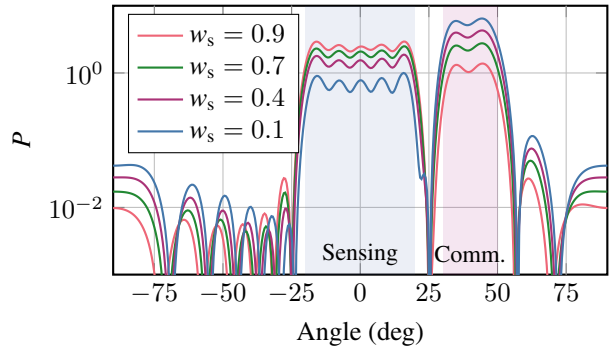

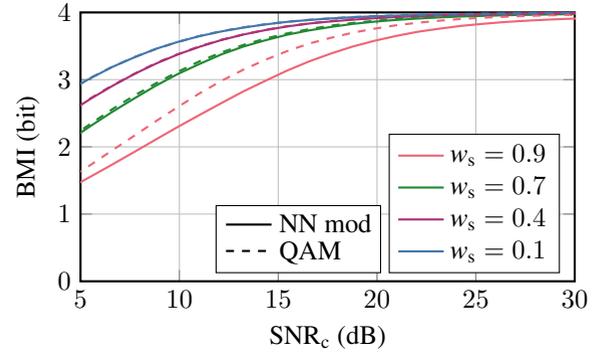
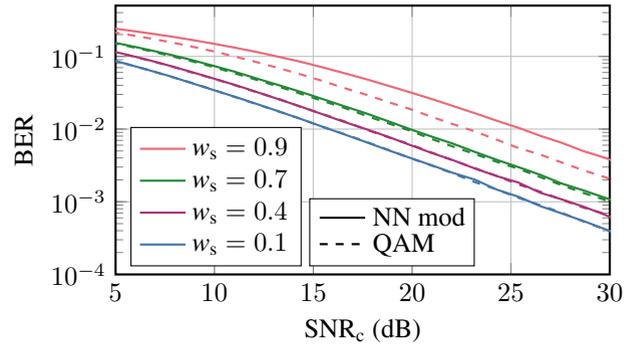
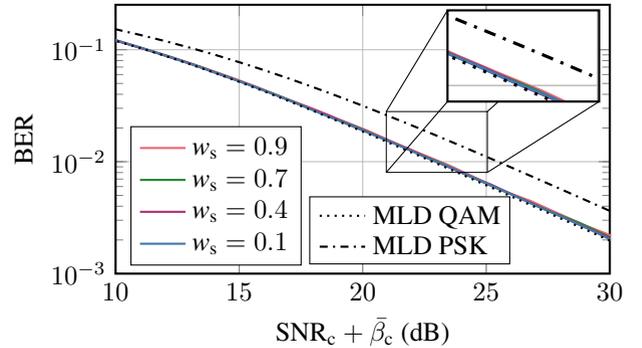
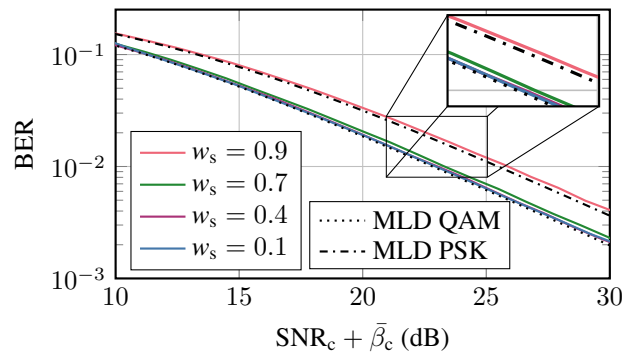
\begin{figure}
\begin{subfigure}[c]{\columnwidth}
\hspace{3.75mm}
    	\begin{tikzpicture}
		\begin{axis}[
			xlabel=$\text{SNR}_{\text{c}}$ (dB), ylabel=BMI (bit),
			grid=major,
			legend cell align={left},
			legend pos=south east,
			xmin=5,xmax=30,
			ymin=0,ymax=4,
			axis line style=thick,
                width=0.95\columnwidth,
                height=0.6\columnwidth,
                legend style={name=leg},
			]
                \addlegendimage{thick, solid, cb-2} \addlegendentry{$w_{\text{s}}=0.9$}
			\addlegendimage{thick, solid, cb-3} \addlegendentry{$w_{\text{s}}=0.7$}
                \addlegendimage{thick, solid, cb-6} \addlegendentry{$w_{\text{s}}=0.4$}
                \addlegendimage{thick, solid, cb-1} \addlegendentry{$w_{\text{s}}=0.1$}
   
			\addplot[mark=none, color=cb-2, thick] table [x expr= 10*log10(\thisrow{SNR}), y = NN0.9]
			{figures/level/NN_loss/SNRsweep_BMI.txt};
                \addplot[mark=none, color=cb-3, thick] table [x expr= 10*log10(\thisrow{SNR}), y = NN0.7]
			{figures/level/NN_loss/SNRsweep_BMI.txt};
                \addplot[mark=none, color=cb-6, thick] table [x expr= 10*log10(\thisrow{SNR}), y = NN0.4]
			{figures/level/NN_loss/SNRsweep_BMI.txt};
                \addplot[mark=none, color=cb-1, thick] table [x expr= 10*log10(\thisrow{SNR}), y = NN0.1]
			{figures/level/NN_loss/SNRsweep_BMI.txt};

                \addplot[mark=none, color=cb-2, dashed, thick] table [x expr= 10*log10(\thisrow{SNR}), y = NN0.9]
			{figures/level/QAM_loss/SNRsweep_BMI.txt};
                \addplot[mark=none, color=cb-3, dashed, thick] table [x expr= 10*log10(\thisrow{SNR}), y = NN0.7]
			{figures/level/QAM_loss/SNRsweep_BMI.txt};
                \addplot[mark=none, color=cb-6, dashed, thick] table [x expr= 10*log10(\thisrow{SNR}), y = NN0.4]
			{figures/level/QAM_loss/SNRsweep_BMI.txt};
                \addplot[mark=none, color=cb-1, dashed, thick] table [x expr= 10*log10(\thisrow{SNR}), y = NN0.1]
			{figures/level/QAM_loss/SNRsweep_BMI.txt};
		\end{axis}
  \node [draw,fill=white, right=-1mm, anchor=south east] at (leg.south west) {\shortstack[l]{
            \ref{Solid} NN mod\\
            \ref{Dashed} QAM}};
	\end{tikzpicture}
\subcaption{\ac{BMI} after training \ac{NN} modulator vs QAM for different $w_{\text{s}}$\label{fig:BMI_QAMvstrain}}
\end{subfigure}\\
\begin{subfigure}[c]{\columnwidth}
\vspace{4mm}
	\begin{tikzpicture}
		\begin{semilogyaxis}[
			xlabel=$\text{SNR}_{\text{c}}$ (dB), ylabel=BER,
			grid=major,
			legend cell align={left},
			legend pos=south west,
			xmin=5,xmax=30,
			ymin=0.0001,%
			axis line style=thick,
                width=0.95\columnwidth,
                height=0.6\columnwidth,
                legend style={name=leg},
			]
                
                \addlegendentry{$w_{\text{s}}=0.9$}
			\addplot[mark=none, color=cb-2, thick] table [x expr= 10*log10(\thisrow{SNR}), y = NN0.9]
			{figures/level/NN_loss/SNRsweep_BER.txt};
                \addlegendentry{$w_{\text{s}}=0.7$}
                \addplot[mark=none, color=cb-3, thick] table [x expr= 10*log10(\thisrow{SNR}), y = NN0.7]
			{figures/level/NN_loss/SNRsweep_BER.txt};
                \addlegendentry{$w_{\text{s}}=0.4$}
                \addplot[mark=none, color=cb-6, thick] table [x expr= 10*log10(\thisrow{SNR}), y = NN0.4]
			{figures/level/NN_loss/SNRsweep_BER.txt};
                \addlegendentry{$w_{\text{s}}=0.1$}
                \addplot[mark=none, color=cb-1, thick] table [x expr= 10*log10(\thisrow{SNR}), y = NN0.1]
			{figures/level/NN_loss/SNRsweep_BER.txt};
                \addplot[mark=none, color=cb-2, dashed, thick] table [x expr= 10*log10(\thisrow{SNR}), y = NN0.9]
			{figures/level/QAM_loss/SNRsweep_BER.txt};
                \addplot[mark=none, color=cb-3, dashed, thick] table [x expr= 10*log10(\thisrow{SNR}), y = NN0.7]
			{figures/level/QAM_loss/SNRsweep_BER.txt};
                \addplot[mark=none, color=cb-6, dashed, thick] table [x expr= 10*log10(\thisrow{SNR}), y = NN0.4]
			{figures/level/QAM_loss/SNRsweep_BER.txt};
                \addplot[mark=none, color=cb-1, dashed, thick] table [x expr= 10*log10(\thisrow{SNR}), y = NN0.1]
			{figures/level/QAM_loss/SNRsweep_BER.txt};
			
		\addplot[label=l1, draw= none, thick] coordinates {(-1,1.5)};
            \label{Solid}
             \addplot[label=dashed, draw= none, thick, dashed] coordinates {(-1,2)};
            \label{Dashed}
\end{semilogyaxis}
\node [draw,fill=white, right=1mm, anchor=south west] at (leg.south east) {\shortstack[l]{
            \ref{Solid} NN mod\\
            \ref{Dashed} QAM}};
	\end{tikzpicture}
\subcaption{BER after training NN modulator vs QAM\label{fig:BER}}
\end{subfigure}
\begin{subfigure}[b]{\columnwidth}
\vspace{4mm}
	\begin{tikzpicture}[spy using outlines={rectangle, magnification=1.5}]
		\begin{semilogyaxis}[
			xlabel=$\text{SNR}_{\text{c}}+\bar{\beta_{\text{c}}}$ (dB), ylabel=BER,
			grid=major,
			legend cell align={left},
			legend pos=south west,
			xmin=10,xmax=30,
			ymin=0.001,%
			axis line style=thick,
                width=0.95\columnwidth,
                height=0.6\columnwidth,
			]
                \addlegendentry{$w_{\text{s}}=0.9$}
                \addlegendentry{$w_{\text{s}}=0.7$}
                \addlegendentry{$w_{\text{s}}=0.4$}
                \addlegendentry{$w_{\text{s}}=0.1$}

                \addplot[mark=none, color=cb-2, thick] table [x expr= 10*log10(\thisrow{SNR}*1.0755205154418945), y = NN0.9]
			{figures/level/QAM_loss/SNRsweep_BER.txt};
                \addplot[mark=none, color=cb-3, thick] table [x expr= 10*log10(\thisrow{SNR}*2.1850438117980957), y = NN0.7]
			{figures/level/QAM_loss/SNRsweep_BER.txt};
                \addplot[mark=none, color=cb-6, thick] table [x expr= 10*log10(\thisrow{SNR}*3.435122013092041), y = NN0.4]
			{figures/level/QAM_loss/SNRsweep_BER.txt};
                \addplot[mark=none, color=cb-1, thick] table [x expr= 10*log10(\thisrow{SNR}*5.201223850250244), y = NN0.1]
			{figures/level/QAM_loss/SNRsweep_BER.txt};

                \addplot[mark=none, color=black, thick, dotted] table [x = esno, y = ber]
			{figures/theober_qam.tex};
                \addplot[mark=none, color=black, thick,dashdotted] table [x = esno, y = ber]
			{figures/theober_psk.tex};
                
                \coordinate (spypoint) at (axis cs:23,0.015);
			\coordinate (spyviewer) at (axis cs:26.5,0.088);	
			\draw [fill=white] ($(spyviewer)+(1cm,0.6cm)$) rectangle ($(spyviewer)-(1cm,0.6cm)$);
			\spy[width=2cm,height=1.2cm, thin, spy connection path={  \draw(tikzspyonnode.south west) -- (tikzspyinnode.south west);\draw (tikzspyonnode.south east) -- (tikzspyinnode.south east);
			\draw (tikzspyonnode.north west) -- (tikzspyinnode.north west);\draw (tikzspyonnode.north east) -- (intersection of  tikzspyinnode.north east--tikzspyonnode.north east and tikzspyinnode.south east--tikzspyinnode.south west);
			;}] on (spypoint) in node at (spyviewer);
		\addplot[draw= none, thick, dotted] coordinates {(2,1.5)};
            \label{Dotted}
             \addplot[ draw= none, thick, dashdotted] coordinates {(2,2)};
            \label{Dashdotted}
\end{semilogyaxis}
\node [draw,fill=white, right=1mm, anchor=south west,yshift=0.15mm] at (leg.south east) {\shortstack[l]{
            \ref{Dotted} MLD QAM\\
            \ref{Dashdotted} MLD PSK}};
	\end{tikzpicture}
	\subcaption{Trained system using QAM compared to \ac{MLD} for QAM and PSK}
	\label{fig:BERcorr2}
\vspace{4mm}
\end{subfigure}\\
\begin{subfigure}[b]{\columnwidth}
	\begin{tikzpicture}[spy using outlines={rectangle, magnification=1.5}]
		\begin{semilogyaxis}[
			xlabel=$\text{SNR}_{\text{c}}+\bar{\beta_{\text{c}}}$ (dB), ylabel=BER,
			grid=major,
			legend cell align={left},
			legend pos=south west,
			xmin=10,xmax=30,
			ymin=0.001,%
			axis line style=thick,
                width=0.95\columnwidth,
                height=0.6\columnwidth,
			]
                \addlegendentry{$w_{\text{s}}=0.9$};
			\addplot[mark=none, color=cb-2, thick] table [x expr= 10*log10(\thisrow{SNR}*1.0776435136795044), y = NN0.9]
			{figures/level/NN_loss/SNRsweep_BER.txt};
                \addlegendentry{$w_{\text{s}}=0.7$};
                \addplot[mark=none, color=cb-3, thick] table [x expr= 10*log10(\thisrow{SNR}*2.2026209831237793), y = NN0.7]
			{figures/level/NN_loss/SNRsweep_BER.txt};
                \addlegendentry{$w_{\text{s}}=0.4$};
                \addplot[mark=none, color=cb-6, thick] table [x expr= 10*log10(\thisrow{SNR}*3.44679594039917), y = NN0.4]
			{figures/level/NN_loss/SNRsweep_BER.txt};
                \addlegendentry{$w_{\text{s}}=0.1$};
                \addplot[mark=none, color=cb-1, thick] table [x expr= 10*log10(\thisrow{SNR}*5.187615871429443), y = NN0.1]
			{figures/level/NN_loss/SNRsweep_BER.txt};

                \addplot[mark=none, color=black, thick, dotted] table [x = esno, y = ber]
			{figures/theober_qam.tex};
                \addplot[mark=none, color=black, thick,dashdotted] table [x = esno, y = ber]
			{figures/theober_psk.tex};
                
                \coordinate (spypoint) at (axis cs:23,0.015);
			\coordinate (spyviewer) at (axis cs:26.5,0.088);	
			\draw [fill=white] ($(spyviewer)+(1cm,0.6cm)$) rectangle ($(spyviewer)-(1cm,0.6cm)$);
			\spy[width=2cm,height=1.2cm, thin, spy connection path={  \draw(tikzspyonnode.south west) -- (tikzspyinnode.south west);\draw (tikzspyonnode.south east) -- (tikzspyinnode.south east);
			\draw (tikzspyonnode.north west) -- (tikzspyinnode.north west);\draw (tikzspyonnode.north east) -- (intersection of  tikzspyinnode.north east--tikzspyonnode.north east and tikzspyinnode.south east--tikzspyinnode.south west);
			;}] on (spypoint) in node at (spyviewer);
		
\end{semilogyaxis}
\node [draw,fill=white, right=1mm, anchor=south west,yshift=0.15mm] at (leg.south east) {\shortstack[l]{
            \ref{Dotted} MLD QAM\\
            \ref{Dashdotted} MLD PSK}};
	\end{tikzpicture}
	\subcaption{Trained system using shaped constellation compared to \ac{MLD} for QAM and PSK}
 \vspace{4mm}
	\label{fig:BERcorr}
\end{subfigure}
\caption{Comparison of communication performance of systems using QAM or constellation shaping}
\end{figure}

\begin{figure}
\pgfplotsset{
colormap={custom}{
samples of colormap={
10 of viridis,
target pos={0,100,400,600,700,750,850,875,900,1000},
}}}
\begin{subfigure}[t]{0.32\columnwidth}
\begin{tikzpicture}
\footnotesize
		\begin{axis}[
			xlabel=$\Re\{\vect{x}\}$, ylabel=$\Im\{\vect{x}\}$,
			grid=major,
                width=1.2\columnwidth,
                height=1.2\columnwidth,
			legend cell align={left},
			legend pos=north east,
			xmin=-1.5,xmax=1.5,
			ymin=-1.5,ymax=1.5,
			axis line style=thick,
                title={$w_{\text{s}}=0.1$},
                x label style={at={(axis description cs:0.5,-0.15)},anchor=north},
                y label style={at={(axis description cs:-0.15,.5)},anchor=south},
			]
                \addplot [index of colormap={1 of custom}, mark=*, only marks, mark size=1pt] table [x expr = \thisrow{1.0I}*cos(-14.8)+\thisrow{1.0Q}*sin(-14.8), y expr = -\thisrow{1.0I}*sin(-14.8)+\thisrow{1.0Q}*cos(-14.8)] {figures/level/ws_fine/10.txt};
		\end{axis}
\end{tikzpicture}
\end{subfigure}
\begin{subfigure}[t]{0.32\columnwidth}
\begin{tikzpicture}
\footnotesize
		\begin{axis}[
			xlabel=$\Re\{\vect{x}\}$, ylabel=$\Im\{\vect{x}\}$,
			grid=major,
                width=1.2\columnwidth,
                height=1.2\columnwidth,
			legend cell align={left},
			legend pos=north east,
			xmin=-1.5,xmax=1.5,
			ymin=-1.5,ymax=1.5,
			axis line style=thick,
                title={$w_{\text{s}}=0.7$},
                x label style={at={(axis description cs:0.5,-0.15)},anchor=north},
                y label style={at={(axis description cs:-0.15,.5)},anchor=south},
			]
                \addplot [index of colormap={4 of custom}, mark=*, only marks, mark size=1pt] table [x expr = \thisrow{1.0I}*cos(-14.8)+\thisrow{1.0Q}*sin(-14.8), y expr = -\thisrow{1.0I}*sin(-14.8)+\thisrow{1.0Q}*cos(-14.8)] {figures/level/ws_fine/70.txt};
		\end{axis}
\end{tikzpicture}
    
\end{subfigure}
\begin{subfigure}[t]{0.32\columnwidth}
\begin{tikzpicture}
 \footnotesize
            \begin{axis}[
			xlabel=$\Re\{\vect{x}\}$, ylabel=$\Im\{\vect{x}\}$,
			grid=major,
                width=1.2\columnwidth,
                height=1.2\columnwidth,
			legend cell align={left},
			legend pos=north east,
			xmin=-1.5,xmax=1.5,
			ymin=-1.5,ymax=1.5,
			axis line style=thick,
                title={$w_{\text{s}}=0.9$},
                x label style={at={(axis description cs:0.5,-0.15)},anchor=north},
                y label style={at={(axis description cs:-0.15,.5)},anchor=south},
			]
                \addplot [index of colormap={8 of custom}, mark=*, only marks, mark size=1pt] table [x expr = \thisrow{1.0I}*cos(-14.8)+\thisrow{1.0Q}*sin(-14.8), y expr = -\thisrow{1.0I}*sin(-14.8)+\thisrow{1.0Q}*cos(-14.8)] {figures/level/ws_fine/90.txt};
		\end{axis}
	\end{tikzpicture}
 \end{subfigure}\\
 \begin{subfigure}[b]{1\columnwidth}
 \centering
	\begin{tikzpicture}
		\begin{axis}[
			xlabel=$\Re\{\vect{x}\}$, ylabel=$\Im\{\vect{x}\}$,
			grid=major,
            width=0.75\columnwidth,
            height=0.75\columnwidth,
			legend cell align={left},
			legend pos=north east,
			xmin=-1.5,xmax=1.5,
			ymin=-1.5,ymax=1.5,
			axis line style=thick,
                x label style={at={(axis description cs:0.5,-0.15)},anchor=north},
                y label style={at={(axis description cs:-0.15,.5)},anchor=south},
                colorbar,
                colormap name=custom,
                point meta max=1, 
                point meta min=0,
                cycle list={[samples of colormap={100 of custom}]},
                colorbar style={
                    ylabel=$w_{\text{s}}$,
                    ylabel style={rotate=-90},
                    },
			]
            \foreach \n in {1,2,...,100}{
            \addplot+ [
            only marks,
            mark size=1pt,
        ] table [x expr = \thisrow{1.0I}*cos(-14.8)+\thisrow{1.0Q}*sin(-14.8), y expr = -\thisrow{1.0I}*sin(-14.8)+\thisrow{1.0Q}*cos(-14.8)] {figures/level/ws_fine/\n.txt};
        };
		\end{axis}
	\end{tikzpicture}

\end{subfigure}
\caption{Constellation evolution with $w_{\text{s}}$ for $\text{SNR}_{\text{c}}=20$\,dB}
\label{fig:constellation_evolution}
\end{figure}
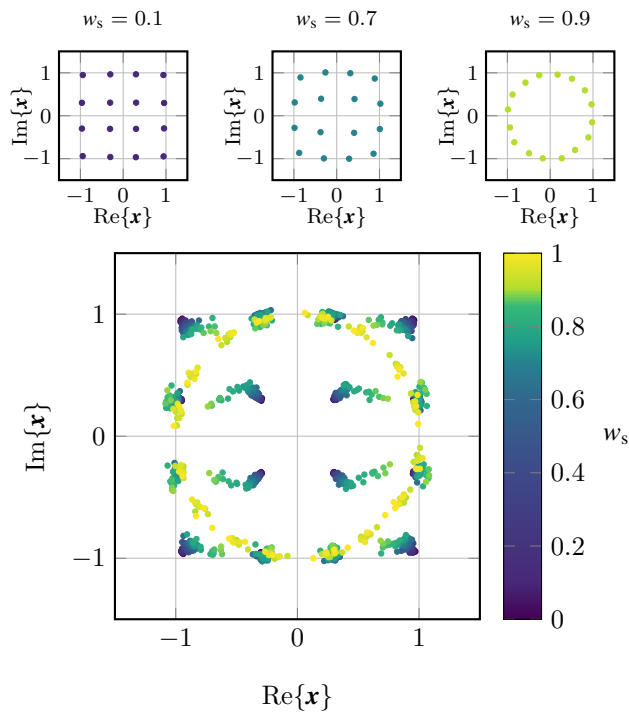

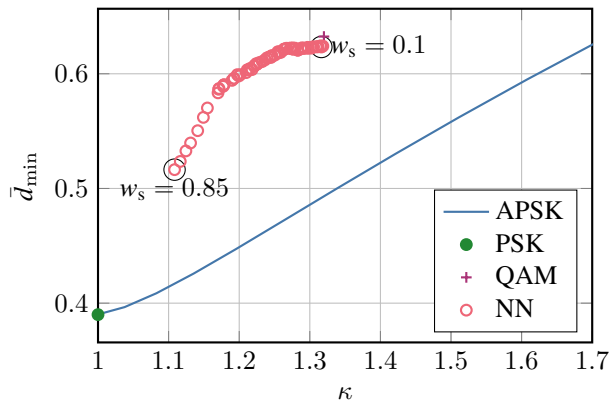
\begin{figure}
    \hspace{3mm}
    \begin{tikzpicture}
		\begin{axis}[
			ylabel=$\bar{d}_{\min}$, 
                xlabel={$\kappa$},
			grid=major,
			legend entries={APSK, PSK, QAM, NN},%
			legend cell align={left},
			legend pos=south east,
			xmin=1,xmax=1.7,
			axis line style=thick,
			tick label style={/pgf/number format/fixed},
                width=0.95\columnwidth,
                height=0.7\columnwidth,
			]
		\addplot[color=cb-1, thick] table [x = kappa, y expr= \thisrow{dmin}]
			{figures/kdminAPSK.txt};
            \addplot[only marks,color=cb-3, mark=*,thick] coordinates {(1, 0.39)};
            \addplot[only marks,color=cb-6, mark=+,thick] coordinates {(528/400, 2/sqrt{10})};
             \addplot[mark=o, only marks, color=cb-2, thick] table [x = kappa, y expr = \thisrow{dmin}]
			{figures/kdmin_mean.txt};

            \draw [black] (axis cs:1.108436979375836628, 0.5163983573400824367) circle (4pt) node [below] {$w_{\text{s}}=0.85$};
            \draw [black] (axis cs:1.316375085838331405e+00, 6.233696616723949369e-01) circle (4pt) node [right] {$w_{\text{s}}=0.1$};
		\end{axis}	
	\end{tikzpicture}
    \caption{Communication performance evaluated based on the mean minimum distance between constellation points and the kurtosis $\kappa$}
    \label{fig:kurtosis}
\end{figure}

\subsection{Beamforming Results}
In Fig.~\ref{fig:ws_comp}, we show the effect of the trade-off parameter $w_{\text{s}}$ on the power radiated to the different areas of interest by summing over the radiated power in the different areas of interest and normalizing with the total power radiated. The power distribution for a system using \ac{QAM} is identical to a system with a shaped constellation.
At $w_{\text{s}}=0.5$, the power is almost equally distributed to the sensing and the communication functionality. For $w_{\text{s}}<0.2$ and $w_{\text{s}}>0.8$, the power radiated toward the less favored function decreases more rapidly. The sum of the radiated power in both areas of interest increases slightly with increasing $w_{\text{s}}$, but approximately $10\%$ of the total power is radiated outside of our areas of interest. 
In Fig.~\ref{fig:ws_comp_fine}, the beam patterns for selected values of $w_{\text{s}}$ are shown.
We radiate more power outside our area of interest while prioritizing communication, which is caused by higher side lobes, particularly at an angle of $\pm 90^{\circ}$. We also observe a slight widening of the communication beam for decreasing $w_{\text{s}}$, which contributes to higher emissions outside our areas of interest.

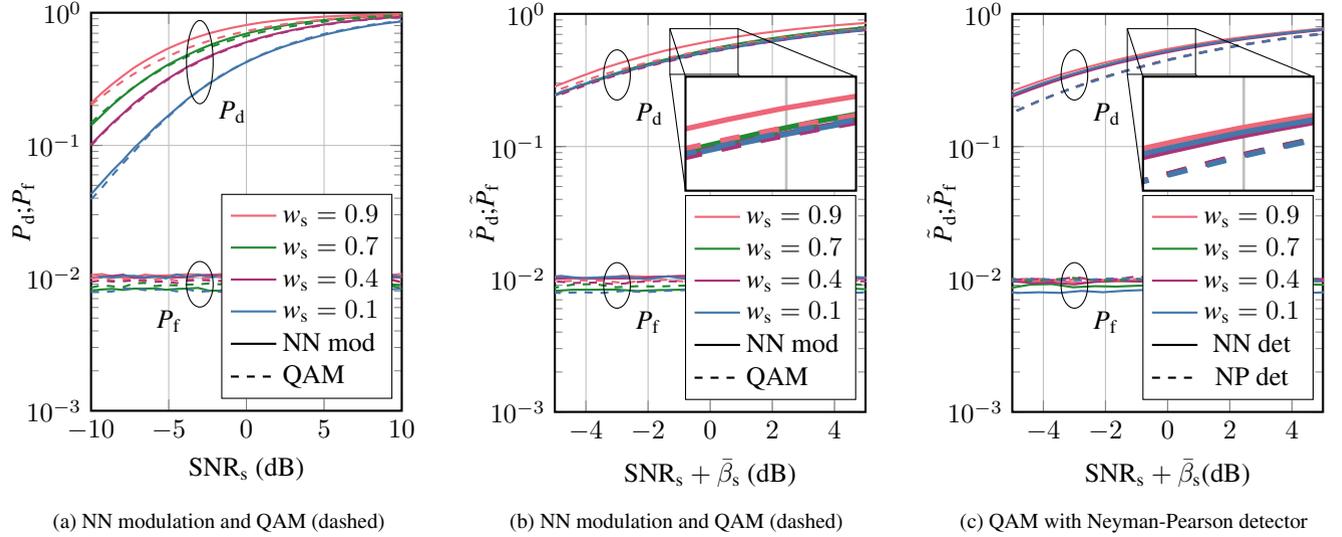
\begin{figure*}
\begin{subfigure}[b]{0.32\textwidth}
	\begin{tikzpicture}
		\begin{semilogyaxis}[
			xlabel=$\text{SNR}_{\text{s}}$ (dB),
                ylabel={$P_{\text{d}}$;$P_{\text{f}}$},
			grid=major,
			legend cell align={left},
			legend pos=south east,
			xmin=-10,xmax=10,
			ymin=0.001,ymax=1,
			axis line style=thick,
			tick label style={/pgf/number format/fixed},
                width=\columnwidth,
                height=1.2\columnwidth,
                y label style={at={(-0.152,0.5)}},
			]
			\addplot[mark=none, color=cb-2, thick] table [x expr= 10*log10(\thisrow{SNR}), y = detect_prob_11]
			{figures/level/NN_loss/SNRsweep_Pd.txt};
			\addplot[mark=none, color=cb-3, thick] table [x expr= 10*log10(\thisrow{SNR}), y = detect_prob_9]
			{figures/level/NN_loss/SNRsweep_Pd.txt};
                \addplot[mark=none, color=cb-6, thick] table [x expr= 10*log10(\thisrow{SNR}), y = detect_prob_6]
			{figures/level/NN_loss/SNRsweep_Pd.txt};
                \addplot[mark=none, color=cb-1, thick] table [x expr= 10*log10(\thisrow{SNR}), y = detect_prob_2]
			{figures/level/NN_loss/SNRsweep_Pd.txt};

                \addlegendentry{$w_{\text{s}}=0.9$}
                \addlegendentry{$w_{\text{s}}=0.7$}
                \addlegendentry{$w_{\text{s}}=0.4$}
                \addlegendentry{$w_{\text{s}}=0.1$}
                \addlegendimage{thick, solid, black} \addlegendentry{NN mod}
			\addlegendimage{thick, dashed, black} \addlegendentry{QAM}
                
                \addplot[mark=none, color=cb-2, dashed, thick] table [x expr= 10*log10(\thisrow{SNR}), y = detect_prob_11]
			{figures/level/QAM_loss/SNRsweep_Pd.txt};
			\addplot[mark=none, color=cb-3, dashed, thick] table [x expr= 10*log10(\thisrow{SNR}), y = detect_prob_9]
			{figures/level/QAM_loss/SNRsweep_Pd.txt};
                \addplot[mark=none, color=cb-6, dashed, thick] table [x expr= 10*log10(\thisrow{SNR}), y = detect_prob_6]
			{figures/level/QAM_loss/SNRsweep_Pd.txt};
                \addplot[mark=none, color=cb-1, dashed, thick] table [x expr= 10*log10(\thisrow{SNR}), y = detect_prob_2]
			{figures/level/QAM_loss/SNRsweep_Pd.txt};
        
                \addplot[mark=none, color=cb-2, thick] table [x expr= 10*log10(\thisrow{SNR}), y = false_alarm_11]
			{figures/level/NN_loss/SNRsweep_Pd.txt};
                \addplot[mark=none, color=cb-3, thick] table [x expr= 10*log10(\thisrow{SNR}), y = false_alarm_9]
			{figures/level/NN_loss/SNRsweep_Pd.txt};
                \addplot[mark=none, color=cb-6, thick] table [x expr= 10*log10(\thisrow{SNR}), y = false_alarm_6]
    		{figures/level/NN_loss/SNRsweep_Pd.txt};
                \addplot[mark=none, color=cb-1, thick] table [x expr= 10*log10(\thisrow{SNR}), y = false_alarm_2]
			{figures/level/NN_loss/SNRsweep_Pd.txt};

                \addplot[mark=none, color=cb-2, dashed, thick] table [x expr= 10*log10(\thisrow{SNR}), y = false_alarm_11]
			{figures/level/QAM_loss/SNRsweep_Pd.txt};
                \addplot[mark=none, color=cb-3, dashed, thick] table [x expr= 10*log10(\thisrow{SNR}), y = false_alarm_9]
			{figures/level/QAM_loss/SNRsweep_Pd.txt};
                \addplot[mark=none, color=cb-6, dashed, thick] table [x expr= 10*log10(\thisrow{SNR}), y = false_alarm_6]
    		{figures/level/QAM_loss/SNRsweep_Pd.txt};
                \addplot[mark=none, color=cb-1, dashed, thick] table [x expr= 10*log10(\thisrow{SNR}), y = false_alarm_2]
			{figures/level/QAM_loss/SNRsweep_Pd.txt};

                \node (xd) at (axis cs:-3,0.6) {};
                \node (yd) at (axis cs:-3,0.3) {};
                \node (xf) at (axis cs:-3,0.01008) {};
                \node (yf) at (axis cs:-3,0.008) {};

		\end{semilogyaxis}
            \node (C1)[draw,ellipse,fit=(yd) (xd),minimum width=0.1cm,inner sep=0pt, label=south east:{$P_{\text{d}}$},] {};
            \node (C2)[draw,ellipse,fit=(yf) (xf),minimum width=0.1cm,inner sep=0pt, label=south west:{$P_{\text{f}}$},] {};

	\end{tikzpicture}
        \vspace{0.075cm}
	\subcaption{NN modulation and QAM (dashed)}
	\label{fig:detection_SNR_NNQAM}
\end{subfigure}
\hspace{2mm}
\begin{subfigure}[b]{0.32\textwidth}
	\begin{tikzpicture}[spy using outlines={rectangle, magnification=2.5}]
		\begin{semilogyaxis}[
			xlabel=$\text{SNR}_{\text{s}}+\bar{\beta}_{\text{s}}$ (dB), 
                ylabel={$\tilde{P}_{\text{d}}$;$\tilde{P}_{\text{f}}$},
			grid=major,
			legend cell align={left},
			legend pos=south east,
			xmin=-5,xmax=5,
			ymin=0.001,ymax=1,
			axis line style=thick,
			tick label style={/pgf/number format/fixed},
                width=\columnwidth,
                height=1.2\columnwidth,
                y label style={at={(-0.152,0.5)}},
			]
			\addplot[mark=none, color=cb-2, thick] table [x expr= 10*log10(\thisrow{SNR}*188.92117309570312/80), y = detect_prob_11]
			{figures/level/NN_loss/SNRsweep_Pd.txt};
			\addplot[mark=none, color=cb-3, thick] table [x expr= 10*log10(\thisrow{SNR}*152.74331665039062/80), y = detect_prob_9]
			{figures/level/NN_loss/SNRsweep_Pd.txt};
                \addplot[mark=none, color=cb-6, thick] table [x expr= 10*log10(\thisrow{SNR}*111.8768539428711/80), y = detect_prob_6]
			{figures/level/NN_loss/SNRsweep_Pd.txt};
                \addplot[mark=none, color=cb-1, thick] table [x expr= 10*log10(\thisrow{SNR}*54.043521881103516/80), y = detect_prob_2]
			{figures/level/NN_loss/SNRsweep_Pd.txt};

                \addlegendentry{$w_{\text{s}}=0.9$}
                \addlegendentry{$w_{\text{s}}=0.7$}
                \addlegendentry{$w_{\text{s}}=0.4$}
                \addlegendentry{$w_{\text{s}}=0.1$}
                \addlegendimage{thick, solid, black} \addlegendentry{NN mod}
			\addlegendimage{thick, dashed, black} \addlegendentry{QAM}
                
                \addplot[mark=none, color=cb-2, dashed, thick] table [x expr= 10*log10(\thisrow{SNR}*189.646728515625/80), y = detect_prob_11]
			{figures/level/QAM_loss/SNRsweep_Pd.txt};
			\addplot[mark=none, color=cb-3, dashed, thick] table [x expr= 10*log10(\thisrow{SNR}*153.37571716308594/80), y = detect_prob_9]
			{figures/level/QAM_loss/SNRsweep_Pd.txt};
                \addplot[mark=none, color=cb-6, dashed, thick] table [x expr= 10*log10(\thisrow{SNR}*112.33573913574219/80), y = detect_prob_6]
			{figures/level/QAM_loss/SNRsweep_Pd.txt};
                \addplot[mark=none, color=cb-1, dashed, thick] table [x expr= 10*log10(\thisrow{SNR}*53.59685516357422/80), y = detect_prob_2]
			{figures/level/QAM_loss/SNRsweep_Pd.txt};
        
                \addplot[mark=none, color=cb-2, thick] table [x expr= 10*log10(\thisrow{SNR}*188.92117309570312/80), y = false_alarm_11]
			{figures/level/NN_loss/SNRsweep_Pd.txt};
                \addplot[mark=none, color=cb-3, thick] table [x expr= 10*log10(\thisrow{SNR}*152.74331665039062/80), y = false_alarm_9]
			{figures/level/NN_loss/SNRsweep_Pd.txt};
                \addplot[mark=none, color=cb-6, thick] table [x expr= 10*log10(\thisrow{SNR}*111.8768539428711/80), y = false_alarm_6]
    		{figures/level/NN_loss/SNRsweep_Pd.txt};
                \addplot[mark=none, color=cb-1, thick] table [x expr= 10*log10(\thisrow{SNR}*54.043521881103516/80), y = false_alarm_2]
			{figures/level/NN_loss/SNRsweep_Pd.txt};

                \addplot[mark=none, color=cb-2, dashed, thick] table [x expr= 10*log10(\thisrow{SNR}*189.646728515625/80), y = false_alarm_11]
			{figures/level/QAM_loss/SNRsweep_Pd.txt};
                \addplot[mark=none, color=cb-3, dashed, thick] table [x expr= 10*log10(\thisrow{SNR}*153.37571716308594/80), y = false_alarm_9]
			{figures/level/QAM_loss/SNRsweep_Pd.txt};
                \addplot[mark=none, color=cb-6, dashed, thick] table [x expr= 10*log10(\thisrow{SNR}*112.33573913574219/80), y = false_alarm_6]
    		{figures/level/QAM_loss/SNRsweep_Pd.txt};
                \addplot[mark=none, color=cb-1, dashed, thick] table [x expr= 10*log10(\thisrow{SNR}*53.59685516357422/80), y = false_alarm_2]
			{figures/level/QAM_loss/SNRsweep_Pd.txt};

                \node (xd) at (axis cs:-3,0.4) {};
                \node (yd) at (axis cs:-3,0.3) {};
                \node (xf) at (axis cs:-3,0.01008) {};
                \node (yf) at (axis cs:-3,0.008) {};
            \coordinate (spypoint) at (axis cs:-0.2,0.52);
			\coordinate (spyviewer) at (axis cs:1.95,0.125);	
			\fill [fill=white] ($(spyviewer)+(1.125cm,0.75cm)$) rectangle ($(spyviewer)-(1.125cm,0.75cm)$);
			\spy[width=2.25cm,height=1.5cm, spy connection path={  \draw(tikzspyonnode.north west) -- (tikzspyinnode.north west);\draw (tikzspyonnode.north east) -- (tikzspyinnode.north east);
			\draw (tikzspyonnode.south west) -- (tikzspyinnode.south west);\draw (tikzspyonnode.south east) -- (intersection of  tikzspyinnode.south east--tikzspyonnode.south east and tikzspyinnode.north east--tikzspyinnode.north west);
			;}] on (spypoint) in node at (spyviewer);
            
		\end{semilogyaxis}
            \node (C1)[draw,ellipse,fit=(yd) (xd),minimum width=0.1cm,inner sep=0pt, label=south east:{$P_{\text{d}}$},] {};
            \node (C2)[draw,ellipse,fit=(yf) (xf),minimum width=0.1cm,inner sep=0pt, label=south east:{$P_{\text{f}}$},] {};

	\end{tikzpicture}
	\subcaption{NN modulation and QAM (dashed)}
	\label{fig:detection_SNRcorr_NNQAM}
\end{subfigure}
\hspace{2mm}
\begin{subfigure}[b]{0.32\textwidth}
	\begin{tikzpicture}[spy using outlines={rectangle, magnification=2.5}]
		\begin{semilogyaxis}[
			xlabel=$\text{SNR}_{\text{s}} +\bar{\beta}_{\text{s}}$(dB), 
                ylabel={$\tilde{P}_{\text{d}}$;$\tilde{P}_{\text{f}}$},
			grid=major,
			legend pos=south east,
			xmin=-5,xmax=5,
			ymin=0.001,ymax=1,
			axis line style=thick,
			extra x tick style={grid=none,tickwidth=0},
			tick label style={/pgf/number format/fixed},
                width=\columnwidth,
                height=1.2\columnwidth,
                y label style={at={(-0.152,0.5)}},
			]
			\addplot[mark=none, color=cb-2, thick] table [x expr= 10*log10(\thisrow{SNR}*189.646728515625/80), y = detect_prob_11]
			{figures/level/QAM_loss/SNRsweep_Pd.txt};
			\addplot[mark=none, color=cb-3, thick] table [x expr= 10*log10(\thisrow{SNR}*153.37571716308594/80), y = detect_prob_9]
			{figures/level/QAM_loss/SNRsweep_Pd.txt};
                \addplot[mark=none, color=cb-6, thick] table [x expr= 10*log10(\thisrow{SNR}*112.33573913574219/80), y = detect_prob_6]
			{figures/level/QAM_loss/SNRsweep_Pd.txt};
                \addplot[mark=none, color=cb-1, thick] table [x expr= 10*log10(\thisrow{SNR}*53.59685516357422/80), y = detect_prob_2]
			{figures/level/QAM_loss/SNRsweep_Pd.txt};

                \addlegendentry{$w_{\text{s}}=0.9$}
                \addlegendentry{$w_{\text{s}}=0.7$}
                \addlegendentry{$w_{\text{s}}=0.4$}
                \addlegendentry{$w_{\text{s}}=0.1$}
                \addlegendimage{thick, solid, black} \addlegendentry{NN det}
			\addlegendimage{thick, dashed, black} \addlegendentry{NP det}
                
                \addplot[mark=none, color=cb-2, dashed, thick] table [x expr= 10*log10(\thisrow{SNR}*189.646728515625/80), y = detect_prob_bm11]
			{figures/level/QAM_loss/SNRsweep_Pd.txt};
			\addplot[mark=none, color=cb-3, dashed, thick] table [x expr= 10*log10(\thisrow{SNR}*153.37571716308594/80), y = detect_prob_bm9]
			{figures/level/QAM_loss/SNRsweep_Pd.txt};
                \addplot[mark=none, color=cb-6, dashed, thick] table [x expr= 10*log10(\thisrow{SNR}*112.33573913574219/80), y = detect_prob_bm6]
			{figures/level/QAM_loss/SNRsweep_Pd.txt};
                \addplot[mark=none, color=cb-1, dashed, thick] table [x expr= 10*log10(\thisrow{SNR}*53.59685516357422/80), y = detect_prob_bm2]
			{figures/level/QAM_loss/SNRsweep_Pd.txt};
        
                \addplot[mark=none, color=cb-2, thick] table [x expr= 10*log10(\thisrow{SNR}*189.646728515625/80), y = false_alarm_11]
			{figures/level/QAM_loss/SNRsweep_Pd.txt};
                \addplot[mark=none, color=cb-3, thick] table [x expr= 10*log10(\thisrow{SNR}*153.37571716308594/80), y = false_alarm_9]
			{figures/level/QAM_loss/SNRsweep_Pd.txt};
                \addplot[mark=none, color=cb-6, thick] table [x expr= 10*log10(\thisrow{SNR}*112.33573913574219/80), y = false_alarm_6]
    		{figures/level/QAM_loss/SNRsweep_Pd.txt};
                \addplot[mark=none, color=cb-1, thick] table [x expr= 10*log10(\thisrow{SNR}*53.59685516357422/80), y = false_alarm_2]
			{figures/level/QAM_loss/SNRsweep_Pd.txt};

                \addplot[mark=none, color=cb-2, dashed, thick] table [x expr= 10*log10(\thisrow{SNR}*189.646728515625/80), y = false_alarm_bm11]
			{figures/level/QAM_loss/SNRsweep_Pd.txt};
                \addplot[mark=none, color=cb-3, dashed, thick] table [x expr= 10*log10(\thisrow{SNR}*153.37571716308594/80), y = false_alarm_bm9]
			{figures/level/QAM_loss/SNRsweep_Pd.txt};
                \addplot[mark=none, color=cb-6, dashed, thick] table [x expr= 10*log10(\thisrow{SNR}*112.33573913574219/80), y = false_alarm_bm6]
    		{figures/level/QAM_loss/SNRsweep_Pd.txt};
                \addplot[mark=none, color=cb-1, dashed, thick] table [x expr= 10*log10(\thisrow{SNR}*53.59685516357422/80), y = false_alarm_bm2]
			{figures/level/QAM_loss/SNRsweep_Pd.txt};

                \node (xd) at (axis cs:-3,0.4) {};
                \node (yd) at (axis cs:-3,0.3) {};
                \node (xf) at (axis cs:-3,0.01008) {};
                \node (yf) at (axis cs:-3,0.008) {};

                \coordinate (spypoint) at (axis cs:-0.2,0.52);
			\coordinate (spyviewer) at (axis cs:1.95,0.125);
			\fill [fill=white] ($(spyviewer)+(1.125cm,0.75cm)$) rectangle ($(spyviewer)-(1.125cm,0.75cm)$);
			\spy[width=2.25cm,height=1.5cm, thin, spy connection path={  \draw(tikzspyonnode.north west) -- (tikzspyinnode.north west);\draw (tikzspyonnode.north east) -- (tikzspyinnode.north east);
			\draw (tikzspyonnode.south west) -- (tikzspyinnode.south west);\draw (tikzspyonnode.south east) -- (intersection of  tikzspyinnode.south east--tikzspyonnode.south east and tikzspyinnode.north east--tikzspyinnode.north west);
			;}] on (spypoint) in node at (spyviewer);
   
		      \end{semilogyaxis}
        \node (C1)[draw,ellipse,fit=(yd) (xd),minimum width=0.1cm,inner sep=0pt, label=south east:{$P_{\text{d}}$},] {};
\node (C2)[draw,ellipse,fit=(yf) (xf),minimum width=0.1cm,inner sep=0pt, label=south east:{$P_{\text{f}}$},] {};
  \end{tikzpicture}
	\subcaption{QAM with Neyman-Pearson detector}
	\label{fig:detection_SNRcorr_bm}
\end{subfigure}
\caption{Detection probability and false alarm rate for varying \ac{SNR} and single snapshot detection}
\end{figure*}

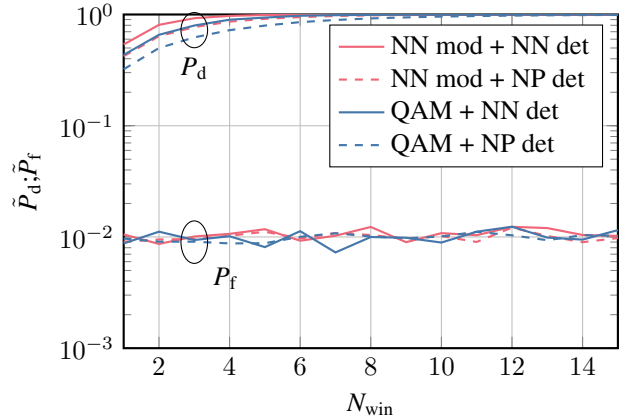
\begin{figure}
    \begin{tikzpicture}
		\begin{semilogyaxis}[
			xlabel=$N_{\text{win}}$, 
                ylabel={$\tilde{P}_{\text d}$;$\tilde{P}_{\text{f}}$},
			grid=major,
			legend entries={NN mod + NN det,NN mod + NP det, QAM + NN det,QAM + NP det}, 
			legend cell align={left},
			legend pos=north east,
			xmin=1,xmax=15,
			ymin=0.001,ymax=1,
			axis line style=thick,
			tick label style={/pgf/number format/fixed},
                width=0.95\columnwidth,
                height=0.7\columnwidth,
			]
			\addplot[mark=none, color=cb-2, thick] table [x = cpr, y = detect_prob_3]
			{figures/level/NN_cpr_sweep_-5/SNRsweep_Pd.txt};
			\addplot[mark=none, color=cb-2, dashed, thick] table [x = cpr, y = detect_prob_bm3]
			{figures/level/NN_cpr_sweep_-5/SNRsweep_Pd.txt};
			 \addplot[mark=none, color=cb-1, thick] table [x = cpr, y = detect_prob_3]
			 {figures/level/QAM_cpr_sweep_-5/SNRsweep_Pd.txt};
		  \addplot[mark=none, color=cb-1, dashed, thick] table [x = cpr, y = detect_prob_bm3]
			 {figures/level/QAM_cpr_sweep_-5/SNRsweep_Pd.txt};
    
    \addplot[mark=none, color=cb-2, thick] table [x = cpr, y = false_alarm_3]
			{figures/level/NN_cpr_sweep_-5/SNRsweep_Pd.txt};
			\addplot[mark=none, color=cb-2, dashed, thick] table [x = cpr, y = false_alarm_bm3]
			{figures/level/NN_cpr_sweep_-5/SNRsweep_Pd.txt};
			 \addplot[mark=none, color=cb-1, thick] table [x = cpr, y = false_alarm_3]
			 {figures/level/QAM_cpr_sweep_-5/SNRsweep_Pd.txt};
		  \addplot[mark=none, color=cb-1, dashed, thick] table [x = cpr, y = false_alarm_bm3]
			 {figures/level/QAM_cpr_sweep_-5/SNRsweep_Pd.txt};
            \node (xd) at (axis cs:3,0.77) {};
            \node (yd) at (axis cs:3,0.68) {};
            \node (xf) at (axis cs:3,0.01008) {};
            \node (yf) at (axis cs:3,0.008) {};
		\end{semilogyaxis}
  \node (C1)[draw,ellipse,fit=(yd) (xd),minimum width=0.1cm,inner sep=0pt, label=south:{$P_{\text{d}}$},] {};
\node (C2)[draw,ellipse,fit=(yf) (xf),minimum width=0.1cm,inner sep=0pt, label=south east:{$P_{\text{f}}$},] {};
	\end{tikzpicture}
	\vspace{-0.3cm}
	\caption{NN modulation and QAM with Neyman-Pearson benchmark, $\text{SNR}_{\text{s}}=-5\,$dB, $w_{\text{s}}=0.9$}
	\label{fig:pdpf_cpr}
	\vspace{-0.4cm}
\end{figure}

\subsection{Communication Results}
In Fig.~\ref{fig:BMI_QAMvstrain}, we show the \ac{BMI} for different $w_{\text{s}}$ values. As $M=16$, the \ac{BMI} cannot exceed $4$ and slowly saturates to this level for high \ac{SNR}. The shaped constellations perform very close to \ac{QAM} for $w_{\text{s}}\geq 0.7$.
Fig.~\ref{fig:BER} displays the \ac{BER} over a range of \ac{SNR} values and different trade-off factors $w_{\text{s}}$.
We observe, as expected, that the \ac{BER} increases for higher $w_{\text{s}}$. Two main effects are responsible for this degradation: Part of the \ac{BER} degradation can be attributed to the shaped constellation diagrams, while the rest is attributed to the beamforming gain toward the communication receiver. We can directly observe the degradation caused by the beamforming gain considering \ac{QAM} in Fig.~\ref{fig:BER}. Since the same \ac{QAM} is used for all $w_{\text{s}}$, the performance gap is caused by the beamforming gain only. 
We can see that \ac{QAM} performs robustly over the different SNR values. The \ac{QAM} results experience an \ac{SNR} offset when varying $w_{\text{s}}$, since the beamforming gain varies for different $w_{\text{s}}$. For $w_{\text{s}}\leq 0.7$, the performance of the shaped constellation and \ac{QAM} are nearly identical.

In Fig.~\ref{fig:BERcorr2}, we can verify that the trained detectors for \ac{QAM} perform very close to the \ac{MLD}. The \ac{SNR} is normalized by beamforming gain $\bar{\beta}_{\text{c}}$ for comparability.

The performance degradation caused by the shaped constellations can be quantified using Fig.~\ref{fig:BERcorr}, where the \ac{SNR} is corrected with the beamforming gain, enabling a raw comparison of the performance caused by changing the constellation. For comparison, we also show the \ac{MLD} results for conventional 16-\ac{QAM} and 16-\ac{PSK} in Fig.~\ref{fig:BERcorr}. We can observe a performance close to the \ac{MLD} for all trained systems. As expected, the performance for low $w_{\text{s}}$ is close the \ac{MLD} for \ac{QAM}, but for $w_{\text{s}}=0.9$ we observe a performance penalty of approximately $0.4\,$dB to the \ac{MLD} for \ac{PSK}. 

Depending on the priority of the communication quality set by $w_{\text{s}}$ and the channel SNR, different constellations are obtained, as shown in Fig.~\ref{fig:constellation_evolution}. In particular, if the sensing priority is very low, the constellation diagram resembles a \ac{QAM}\footnote{If the BER is optimized instead of symbol error rates, \ac{QAM} shows better performance than hexagonal lattices, as effective Gray coding in a hexagonal lattice is not possible}. When the priority of the sensing task increases, the inner symbols are slowly pushed outwards. At very high sensing priority ($w_{\text{s}}\approx1$), the constellation resembles a \ac{PSK}-like constellation, as the transmit power becomes equal over all symbols, resulting in a constant modulus signal. This indicates the importance of the kurtosis or a constant modulus constraint for \ac{JCAS} systems. As the constellation converges to a constant amplitude with increasing $w_{\text{s}}$, communication becomes less reliable since the Euclidean distance between the constellation symbols becomes smaller. 

Finally, in Fig.~\ref{fig:kurtosis}, we show the effect of the constellation kurtosis on the mean minimum distance $\bar{d}_{\min}$ by comparing the geometrically shaped constellations obtained from the \ac{NN} to the \ac{APSK} proposed in Sec.~\ref{sec:kurtosis}, a \ac{QAM} and a pure \ac{PSK}. As expected, a higher kurtosis allows the constellation points to be spaced further apart, resulting in a higher $\bar{d}_{\min}$ and a lower \ac{BER} eventually. The reference \ac{APSK} clearly shows this relationship between the kurtosis and the mean minimum distance $\bar{d}_{\min}$ in Fig.~\ref{fig:kurtosis}.
The $\bar{d}_{\min}$ of the trained constellations is higher than the $\bar{d}_{\min}$ of the \ac{APSK}. With decreasing $w_{\text{s}}$, the kurtosis $\kappa$ and the $\bar{d}_{\min}$ increase, approaching the $\bar{d}_{\min}$ of a \ac{QAM}.
Especially for higher-order modulation formats, the kurtosis $\kappa$ manifests as an additional trade-off parameter for sensing and communication performance.

\subsection{Target Detection Results}
In Fig.~\ref{fig:detection_SNR_NNQAM}, we compare the performance of the trained detectors at varying \ac{SNR}. $\tilde{P}_{\text{f}}$ is approximately constant at $10^{-2}$, as intended by design. The detection rate is very similar for both modulation methods, only for high $w_{\text{s}}$, we can see a slightly better detection rate for the shaped constellation. We converge to $\tilde{P}_{\text{d}}\approx 1$ for all cases. The impact of beamforming is clearly visible, especially for very low $w_{\text{s}}$, where the beamformer barely illuminates the sensing area and $\tilde{P}_{\text{d}}=0.5$ is obtained only at $\text{SNR}_{\text{s}}=2\,$dB. 
In Fig.~\ref{fig:detection_SNRcorr_NNQAM}, we correct for the average beamforming gain and observe that most detectors converge to a very similar performance. 
In Fig.~\ref{fig:detection_SNRcorr_bm} the trained detectors are shown with the  benchmark \ac{NP}-based detectors. All \ac{NN}-based detectors trained for \ac{QAM} perform almost identically and we can observe a reduced detection rate of the \ac{NP}-based power detectors.

We evaluate target detection by comparing the detection rate $\tilde{P}_{\text{d}}$ and the false alarm rate $\tilde{P}_{\text{f}}$ for different observation window lengths $N_{\text{win}}$ for $w_{\text{s}}=0.9$ in Fig.~\ref{fig:pdpf_cpr}. Increasing $N_{\text{win}}$ improves the detection rate, as expected. The false alarm rate remains approximately constant for varying $N_{\text{win}}$.
The \ac{NP} benchmark detector leads to a lower detection rate for both setups, since it only relies on the input power. The trained detector can also take directional information into account.

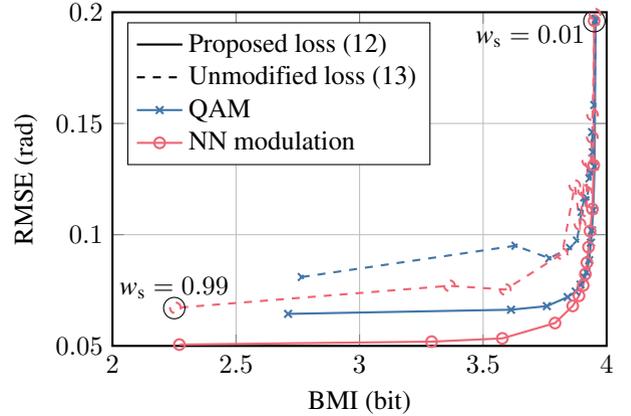
\begin{figure}
	\begin{tikzpicture}
		\begin{axis}[
			xlabel=BMI (bit), ylabel=RMSE (rad),
			grid=major,
			legend entries={QAM,NN}, 
			legend cell align={left},
			legend pos=north west,
			xmin=2,xmax=4,
			ymin=0.05,ymax=0.2,
			axis line style=thick,
                yticklabel style={
                    /pgf/number format/precision=3,
                    /pgf/number format/fixed},
                scaled ticks=false, %
                width=0.95\columnwidth,
                height=0.7\columnwidth,
			]
                \addlegendimage{thick, solid, black} \addlegendentry{Proposed loss~\eqref{eq:loss}}
			\addlegendimage{thick, dashed, black} \addlegendentry{Unmodified loss~\eqref{eq:lossalt}}

            \addplot[mark=x, color=cb-1, thick] table [x = tensor(1.7433)_tensor(96.8739)bmi, y = tensor(1.7433)_tensor(96.8739)rmse]
			{figures/level/QAM_loss/SNRsweep_BMI_vs_rmse.txt};
   \addlegendentry{QAM}
    \addplot[mark=o, color=cb-2, thick] table [x = tensor(1.7433)_tensor(96.8739)bmi, y = tensor(1.7433)_tensor(96.8739)rmse]
			{figures/level/NN_loss/SNRsweep_BMI_vs_rmse.txt};
    \addlegendentry{NN modulation}
   \addplot[mark=x, color=cb-1, thick,dashed] table [x = tensor(1.7433)_tensor(96.8739)bmi, y = tensor(1.7433)_tensor(96.8739)rmse]
			{figures/level/QAM_originloss/SNRsweep_BMI_vs_rmse.txt};
    \addplot[mark=o, color=cb-2, thick,dashed] table [x = tensor(1.7433)_tensor(96.8739)bmi, y = tensor(1.7433)_tensor(96.8739)rmse]
			{figures/level/NN_originloss/SNRsweep_BMI_vs_rmse.txt};
   \draw [black] (axis cs:3.95,0.196) circle (4pt) node [left,yshift=-0.2cm] {$w_{\text{s}}=0.01$};
   \draw [black] (axis cs:2.25,0.067) circle (4pt) node [above] {$w_{\text{s}}=0.99$};
\end{axis}
\end{tikzpicture}
\caption{QAM and \ac{NN} modulator operating at different $w_{\text{s}}$ after training with the modified loss function (solid) and without (dashed). Performance evaluated at $\text{SNR}_{\text{c}}=20.8\,\text{dB}$, $\text{SNR}_{\text{s}}=2.6\,\text{dB}$ and $N_{\text{win}}=1$}
\label{fig:BMI_vs_rmse}
\end{figure}

\begin{figure*}
\begin{subfigure}[t]{0.32\textwidth}
	\begin{tikzpicture}
		\begin{axis}[
			xlabel=$\text{SNR}_{\text{s}}\vphantom{+\bar{\beta}_{\text{s}}}$ (dB), ylabel=Bias (rad),
			grid=major,
			legend cell align={left},
			legend pos=south east,
			xmin=-10,xmax=10,
			ymin=-0.015,ymax=0.015,
			axis line style=thick,
			extra x tick style={grid=none,tickwidth=0},
			ticklabel style={/pgf/number format/fixed},
                scaled ticks=false, %
                width=\columnwidth,
                height=1.2\columnwidth,
                y label style={at={(-0.152,0.5)}},
			]
                \addlegendimage{thick, solid, cb-2} \addlegendentry{$w_{\text{s}}=0.9$}
			\addlegendimage{thick, solid, cb-3} \addlegendentry{$w_{\text{s}}=0.7$}
                \addlegendimage{thick, solid, cb-6} \addlegendentry{$w_{\text{s}}=0.4$}
                \addlegendimage{thick, solid, cb-1} \addlegendentry{$w_{\text{s}}=0.1$}
   
			\addplot[mark=none, color=cb-2, thick] table [x expr= 10*log10(\thisrow{SNR}), y = NN0.9]
			{figures/level/QAM_loss/biases.txt};
			\addplot[mark=none, color=cb-3, thick] table [x expr= 10*log10(\thisrow{SNR}), y = NN0.7]
			{figures/level/QAM_loss/biases.txt};
   \addplot[mark=none, color=cb-6, thick] table [x expr= 10*log10(\thisrow{SNR}), y = NN0.4]
			{figures/level/QAM_loss/biases.txt};
   \addplot[mark=none, color=cb-1, thick] table [x expr= 10*log10(\thisrow{SNR}), y = NN0.1]
			{figures/level/QAM_loss/biases.txt};
		\end{axis}
	\end{tikzpicture}
	\subcaption{\ac{AoA} estimation bias for \ac{QAM} modulation.\\ \phantom{Achieable area is marked with something}}
	\label{fig:bias}
\end{subfigure}
\hspace{2mm}
\begin{subfigure}[t]{0.32\textwidth}
	\begin{tikzpicture}
		\begin{axis}[
			xlabel=$\text{SNR}_{\text{s}}\vphantom{+\bar{\beta}_{\text{s}}}$ (dB), ylabel=RMSE (rad),
			grid=major,
			legend cell align={left},
			legend pos=north east,
			xmin=-10,xmax=10,
			ymin=0,ymax=0.25,
			axis line style=thick,
			extra x tick style={grid=none,tickwidth=0},
			tick label style={/pgf/number format/fixed},
                width=\columnwidth,
                height=1.2\columnwidth,
                y label style={at={(-0.152,0.5)}},
			]
                \addlegendimage{thick, solid, cb-2} \addlegendentry{$w_{\text{s}}=0.9$}
			\addlegendimage{thick, solid, cb-3} \addlegendentry{$w_{\text{s}}=0.7$}
                \addlegendimage{thick, solid, cb-6} \addlegendentry{$w_{\text{s}}=0.4$}
                \addlegendimage{thick, solid, cb-1} \addlegendentry{$w_{\text{s}}=0.1$}
			\addlegendimage{thick, solid, black} \addlegendentry{NN mod}
			\addlegendimage{thick, dashed, black} \addlegendentry{QAM}
   
			\addplot[mark=none, color=cb-2, thick] table [x expr= 10*log10(\thisrow{SNR}), y = rmse_h11]
			{figures/level/NN_loss/SNRsweep_rmse.txt};
			\addplot[mark=none, color=cb-3, thick] table [x expr= 10*log10(\thisrow{SNR}), y = rmse_h9]
			{figures/level/NN_loss/SNRsweep_rmse.txt};
                \addplot[mark=none, color=cb-6, thick] table [x expr= 10*log10(\thisrow{SNR}), y = rmse_h6]
			{figures/level/NN_loss/SNRsweep_rmse.txt};
                \addplot[mark=none, color=cb-1, thick] table [x expr= 10*log10(\thisrow{SNR}), y = rmse_h2]
			{figures/level/NN_loss/SNRsweep_rmse.txt};

                \addplot[mark=none, color=cb-2, dashed, thick] table [x expr= 10*log10(\thisrow{SNR}), y = rmse_h11]
			{figures/level/QAM_loss/SNRsweep_rmse.txt};
			\addplot[mark=none, color=cb-3, dashed, thick] table [x expr= 10*log10(\thisrow{SNR}), y = rmse_h9]
			{figures/level/QAM_loss/SNRsweep_rmse.txt};
                \addplot[mark=none, color=cb-6, dashed, thick] table [x expr= 10*log10(\thisrow{SNR}), y = rmse_h6]
			{figures/level/QAM_loss/SNRsweep_rmse.txt};
                \addplot[mark=none, color=cb-1, dashed, thick] table [x expr= 10*log10(\thisrow{SNR}), y = rmse_h2]
			{figures/level/QAM_loss/SNRsweep_rmse.txt};
		\end{axis}
	\end{tikzpicture}
	\subcaption{\ac{RMSE}\\ \phantom{Achievable area is marked}}
	\label{fig:rmse_SNR}
\end{subfigure}
\hspace{2mm}
\begin{subfigure}[t]{0.32\textwidth}
	\begin{tikzpicture}
		\begin{axis}[
			xlabel=$\text{SNR}_{\text{s}}+\bar{\beta}_{\text{s}}$ (dB), ylabel=RMSE (rad),
			grid=major,
			legend cell align={left},
			legend pos=south west,
			xmin=-5,xmax=5,
			ymin=0.0,ymax=0.15,
			axis line style=thick,
			extra x tick style={grid=none,tickwidth=0},
			tick label style={/pgf/number format/fixed},
                width=\columnwidth,
                height=1.2\columnwidth,
                y label style={at={(-0.152,0.5)}},
			]
                \addlegendimage{thick, solid, cb-2} \addlegendentry{$w_{\text{s}}=0.9$}
			\addlegendimage{thick, solid, cb-3} \addlegendentry{$w_{\text{s}}=0.7$}
                \addlegendimage{thick, solid, cb-6} \addlegendentry{$w_{\text{s}}=0.4$}
                \addlegendimage{thick, solid, cb-1} \addlegendentry{$w_{\text{s}}=0.1$}
			\addlegendimage{thick, solid, black} \addlegendentry{NN mod}
			\addlegendimage{thick, dashed, black} \addlegendentry{QAM}
   
			\addplot[mark=none, color=cb-2, thick] table [x expr= 10*log10(\thisrow{SNR}*188.92117309570312/80), y = rmse_h11]
			{figures/level/NN_loss/SNRsweep_rmse.txt};
			\addplot[mark=none, color=cb-3, thick] table [x expr= 10*log10(\thisrow{SNR}*152.74331665039062/80), y = rmse_h9]
			{figures/level/NN_loss/SNRsweep_rmse.txt};
                \addplot[mark=none, color=cb-6, thick] table [x expr= 10*log10(\thisrow{SNR}*111.8768539428711/80), y = rmse_h6]
			{figures/level/NN_loss/SNRsweep_rmse.txt};
                \addplot[mark=none, color=cb-1, thick] table [x expr= 10*log10(\thisrow{SNR}*54.043521881103516/80), y = rmse_h2]
			{figures/level/NN_loss/SNRsweep_rmse.txt};

                \addplot[mark=none, color=cb-2, dashed, thick] table [x expr= 10*log10(\thisrow{SNR}*189.646728515625/80), y = rmse_h11]
			{figures/level/QAM_loss/SNRsweep_rmse.txt};
			\addplot[mark=none, color=cb-3, dashed, thick] table [x expr= 10*log10(\thisrow{SNR}*153.37571716308594/80), y = rmse_h9]
			{figures/level/QAM_loss/SNRsweep_rmse.txt};
                \addplot[mark=none, color=cb-6, dashed, thick] table [x expr= 10*log10(\thisrow{SNR}*112.33573913574219/80), y = rmse_h6]
			{figures/level/QAM_loss/SNRsweep_rmse.txt};
                \addplot[mark=none, color=cb-1, dashed, thick] table [x expr= 10*log10(\thisrow{SNR}*53.59685516357422/80), y = rmse_h2]
			{figures/level/QAM_loss/SNRsweep_rmse.txt};
        \addplot[name path=crb,color=KITred, thick] coordinates {
        (-5,0.02375523) (-4.5,0.02222424) (-4,0.02080947) (-3.5,0.01949992) (-3,0.01828583) (-2.5,0.01715855) (-2,0.01611037) (-1.5,0.01513445) (-1,0.01422468) (-0.5,0.0133756) (0,0.0125823) (0.5,0.01184041)
 (1,0.01114595) (1.5,0.01049536) (2,0.0098854) (2.5,0.00931315) (3,0.00877593) (3.5,0.00827132) (4,0.00779709)  (4.5,0.0073512) (5,0.00693179)
        };
\path[name path=axis,only marks,mark=none] (axis cs:-5,0.0001) -- (axis cs:5,0.0001);
 
\addplot [
        color=KITred,
        fill=KITred, 
        fill opacity=0.3,
    ]
    fill between[of=crb and axis,
    ]; 
     \node[KITred] (crb) at (axis cs: 2,0.017){CRB};
		\end{axis}
	\end{tikzpicture}
	\subcaption{\ac{RMSE} with \ac{SNR} corrected with $\bar{\beta}_{\text{s}}$. Achievable area is limited by \ac{CRB} in red delimiting the shaded area.}
	\label{fig:rmse_SNRcorr}
\end{subfigure}
\caption{\ac{AoA} estimation results evaluated on different \ac{SNR} for single snapshot sensing}
\vspace{-0.4cm}
\end{figure*}
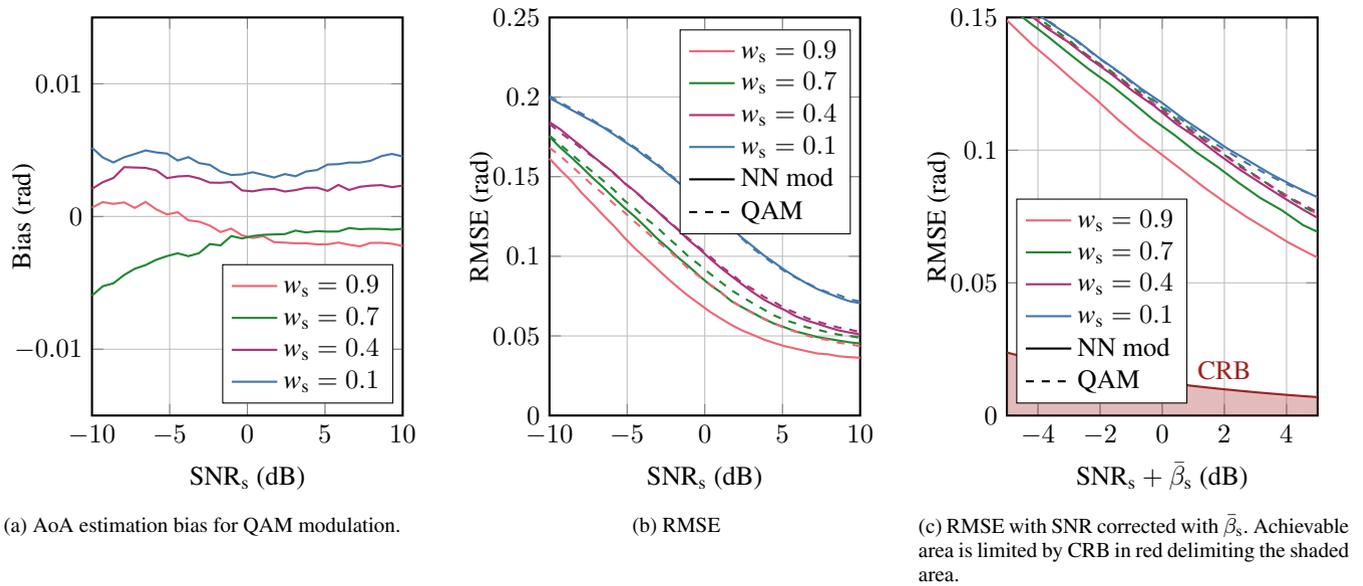

\begin{figure}
	\begin{tikzpicture}
		\begin{semilogyaxis}[
			xlabel=Window length $N_{\text{win}}$, ylabel={RMSE (rad)},
			grid=major,
			legend cell align={left},
			legend pos=south west,
			xmin=1,xmax=15,
			ymin=0.0001,ymax=1,
			axis line style=thick,
			extra x ticks={1},
			extra x tick style={grid=none,tickwidth=0},
			tick label style={/pgf/number format/fixed},
                width=0.95\columnwidth,
                height=0.7\columnwidth,
			]
                \addlegendimage{thick, solid, cb-2} \addlegendentry{NN mod}
			\addlegendimage{thick, solid, cb-1} \addlegendentry{QAM}
			\addlegendimage{thick, solid, black} \addlegendentry{NN estimation}
			\addlegendimage{thick, dashed, black} \addlegendentry{ESPRIT}
			\addplot[mark=none, color=cb-2, thick] table [x = cpr, y = rmse_h3]
{figures/level/NN_cpr_sweep_-5/SNRsweep_rmse.txt};
			\addplot[mark=none, color=cb-2, dashed, thick] table [x = cpr, y = rmse_esprit3]
			{figures/level/NN_cpr_sweep_-5/SNRsweep_rmse.txt};
			 \addplot[mark=none, color=cb-1, thick] table [x = cpr, y = rmse_h3]
			 {figures/level/QAM_cpr_sweep_-5/SNRsweep_rmse.txt};
		  \addplot[mark=none, color=cb-1, dashed, thick] table [x = cpr, y = rmse_esprit3]
			 {figures/level/QAM_cpr_sweep_-5/SNRsweep_rmse.txt};

    \addplot[name path=crb,color=KITred, thick] table [x = N, y = CRB]
{figures/level/NN_cpr_sweep_-5/crb.txt};
\path[name path=axis,only marks,mark=none] (axis cs:0,0.0001) -- (axis cs:15,0.0001);

\addplot [
        color=KITred,
        fill=KITred, 
        fill opacity=0.3,
    ]
    fill between[of=crb and axis,
    ];

  \node[KITred] (crb) at (axis cs: 11,0.0003){CRB};
		\end{semilogyaxis}
  
	\end{tikzpicture}
	\vspace{-0.3cm}
	\caption{\ac{RMSE} of angle estimation compared to ESPRIT algorithm for QAM and NN-based modulation at an \ac{SNR} of $-5\,$dB for the sensing channel with $w_{\text{s}}=0.9$. Achievable area is limited by \ac{CRB}.}
	\label{fig:rmse_cpr}
	\vspace{-0.4cm}
\end{figure}
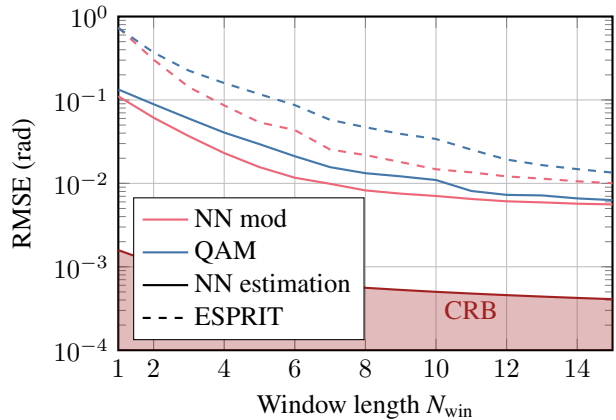

Having reviewed the effects of multi-snapshot estimation and the use of different modulation formats, the presented setups enables us to choose different trade-offs between communication and sensing. Even though we sacrifice some sensing performance through the use of \ac{QAM}, we can significantly improve detection and parameter estimation by collecting multiple samples to perform sensing in scenarios where objects are slow enough to be captured by multiple samples at almost the same position.

\subsection{Angle Estimation Results}
We analyze the effect of the modified angle loss term on the results compared to the unmodified loss given in~\eqref{eq:lossalt} in Fig.~\ref{fig:BMI_vs_rmse}. We optimize the \ac{JCAS} system by choosing different trade-off factors $w_{\text{s}}$ and display the \ac{RMSE} of the angle of estimation as a function of the \ac{BMI} for each value of $w_{\text{s}}$. 
With the original loss function~\eqref{eq:lossalt}, as indicated by the dashed lines, there are performance fluctuations. After training, the estimation error depends mainly on the noise power and $N_{\text{win}}$. Since $N_{\text{win}}$ and $\sigma_{\text{ns}}$ are randomly chosen during training, the loss is influenced by random processes which distort the magnitude of the gradients. Nevertheless, this randomness during training is beneficial for generalization.
During training, we also noticed that trained systems converged to solutions of varying performance. We would anticipate a monotonic behavior of the curves, from high $w_{\text{s}}$ in the left bottom corner of the plot to low $w_{\text{s}}$ in the upper right corner. By modifying the loss function, we keep the randomness in our training data while normalizing the expected magnitude of the gradients. We achieve a gradual performance trade-off as expected and achieve a lower \ac{AoA} estimation \ac{RMSE} than for the unmodified loss while a similar \ac{BMI} is achieved. 
The trained constellation generally achieves better accuracy in estimating the \ac{AoA}, which can be attributed to its smaller constellation kurtosis.%

We investigate the estimation bias of the \ac{AoA} \ac{NN} in Fig.~\ref{fig:bias}. The trained estimators lead to small biases in the order of $10^{-2}$. 
There is not a clear trend of the bias as a function of \ac{SNR} or $w_{\text{s}}$, or a systematic bias.
In Fig.~\ref{fig:rmse_SNR}, we can observe how increasing $w_{\text{s}}$ decreases the angle estimation error. In particular, while the changes of \ac{BER} for $w_{\text{s}}=0.7$ in Fig.~\ref{fig:BER} between trained constellation and \ac{QAM} are not very distinct, there is a gain in sensing performance. From Fig.~\ref{fig:rmse_SNRcorr}, we can observe that for larger $w_{\text{s}}$, the gap of \ac{NN} modulation and \ac{QAM} increases further. Furthermore, as the priority of sensing is very low for $w_{\text{s}}=0.1$, the sensing does not approach the results of the other detectors for \ac{QAM}. We know from Fig.~\ref{fig:ws_comp_fine} that at $w_{\text{s}}=0.1$, the beam towards the sensing targets varies in magnitude in our area of interest more significantly than for higher $w_{\text{s}}$, indicating that the minima in the wider beam lead to a high \ac{RMSE} for targets. 
in Fig.~\ref{fig:rmse_cpr}, we compare different window lengths $N_{\text{win}}$. The trained angle estimators outperform ESPRIT at a raw $\text{SNR}_{\text{s}}=-5\,$dB. At low \ac{SNR}, the proposed method can consistently outperform the \ac{ESPRIT} baseline. The gap between the modulation formats first becomes larger with increasing window length until $N_{\text{win}}\approx 4$ for both modulation methods. Since both modulators lead to the same average output power, a longer observation window reduces the variance of the auto-correlation, under the assumption of a similar statistic for both modulation formats. There is a gap to the \ac{CRB} for all estimators as expected in such a low \ac{SNR} scenario.

\section{Scenario Extension: Multi-User MIMO}
We extend the analyzed scenario to the downlink of multi-user \ac{MIMO}, to demonstrate how spatial efficiency can be addressed with our setup. In this section, we show a proof-of-concept how to extend the system of Fig.~\ref{fig:flowgraphtrain_mono} to $N_{\text{ue}}$ \acp{UE}. The beamforming \ac{NN} needs to be extended
to enable precoding for multiple different positions, therefore we enlarge the output layer to $2K\cdot N_{\text{ue}}$ neurons. The inputs of the precoding \ac{NN} are then $\{\varphi_{\min},\varphi_{\max}, \theta_{\text{ue}1},\ldots ,\theta_{\text{ue}N_{\text{ue}}}\}$ and
\eqref{eq:beam} is replaced by
\begin{align}
    \mat{y} = \mat{v} \cdot \mat{X},
\end{align}
with $\mat{v} \in \mathbb{C}^{K \times N_{\text{ue}}}$ being the precoding matrix and $\mat{X} \in \mathbb{C}^{N_{\text{ue}}\times N_{\text{win}}}$ denoting the transmit symbols for all \acp{UE}.

We need to use a different loss term for communication to address the resource allocation. 
We reformulate the maximization of an $\alpha$-fair utility function or the weighted sum rate as in \cite[(5)]{Castaneda2017} to a loss function as
\begin{align}
&\tilde{L}_{\text{comm}}(\alpha) =\notag\\
&\begin{cases}
-\sum_{n=1}^{N_{\text{ue}}}\log_2 \left(\sum_{i=1}^{\log_2M} \left[1-H(\mathsf{b}_{i,n}||\hat{\mathsf{b}}_{i,n})\right]\right) &, \alpha=1 \\
-\sum_{n=1}^{N_{\text{ue}}}  \frac{\left(\sum_{i=1}^{\log_2M} \left[1-H(\mathsf{b}_{i,n}||\hat{\mathsf{b}}_{i,n})\right]\right)^{1-\alpha}  }{1-\alpha}      &,\alpha \neq 1,
    \end{cases} 
\end{align}
with $\alpha\in \mathbb{R}^+$ trading off the total throughput and fairness between \acp{UE}. $\alpha=0$ results in a sum-rate maximization and $\alpha=1$ yields maximum fairness. Fairness is especially important in scenarios, where the channel capacities between \ac{UE} and the base station vary significantly between \acp{UE} to prevent \acp{UE} with low channel capacity from being allocated no resources at all. In our training setup, $\tilde{L}_{\text{comm}}(\alpha)$ replaces $L_{\text{comm}}$ in \eqref{eq:loss}.

We demonstrate such a system with 2 \acp{UE} whose dominant reflection path is at an angle of departure of $50^{\circ}$ and $70^{\circ}$ respectively. The simulated \ac{SNR} range is the same for both \acp{UE} and we train a respective decoder while Q\ac{PSK} is used for transmission. This setup results in the same total data throughput as the other simulations in this paper. Both \acp{UE} attempt to receive the same number of symbols. For the communication loss $\tilde{L}_{\text{comm}}(\alpha)$ we choose $\alpha=\nobreak1$ and we set $w_{\text{s}}=0.7$.

The effects of the precoding are shown in Fig.~\ref{fig:mimo-beam}. The vertical purple lines mark the direction of the main reflection servicing the \acp{UE}. The average total beam pattern is shown as well as the effects of the precoding vectors for each \ac{UE}. The beam pattern for each \ac{UE} shows a maximum towards its target direction and minimum towards the other \ac{UE}. Both signals contribute to the illumination of the sensing area of interest. At the same \ac{SNR}, the \ac{BER} is similar for both \acp{UE}, as shown in Fig.~\ref{fig:mimo-ber}. The beamforming gain is slightly lower for UE2, therefore we see a slightly higher \ac{BER}. The \ac{BER} is generally lower as in the other presented scenario, as Q\ac{PSK} instead of a higher order modulation is used.

\begin{figure}
	\begin{tikzpicture}
		\begin{semilogyaxis}[
			xlabel=Angle (deg), ylabel=$P$ ,
			legend entries={UE1,UE2, Sum}, 
			legend cell align={left},
			legend pos=south west,
			xmin=-90,xmax=90,
			ymin=0.001,ymax=10,
			axis line style=thick,
                width=0.95\columnwidth,
                height=0.6\columnwidth,
                extra x ticks={-25, 25, 75,-75},
                tick align=inside,
                grid=major,
			]
            \fill[fill=KITblue, fill opacity=0.1] (axis cs: -20,0.001) rectangle (axis cs: 20,10);
            
            \addplot [color=cb-2, thick] table [x expr= {\thisrow{angles}}, y expr = {\thisrow{Ephi0}}]
		{figures/mimo/beampattern_n.txt};
            \addplot [color=cb-3, thick] table [x expr= {\thisrow{angles}}, y expr = {\thisrow{Ephi1}}]
		{figures/mimo/beampattern_n.txt};
            \addplot [color=cb-1, thick] table [x expr= {\thisrow{angles}}, y expr = {\thisrow{sum}}]
		{figures/mimo/beampattern_n.txt};
            \draw [black] (axis cs:0,5) node {\small Sensing};
            \node[KITpurple, rotate=90] at (axis cs:46.5,4.5) {\small UE1};
            \node[KITpurple, rotate=90] at (axis cs:66.5,4.5) {\small UE2};

            \addplot [mark=none, KITpurple,thick] coordinates {(50, 0.00001) (50, 30)};
            \addplot [mark=none, KITpurple,thick] coordinates {(70, 0.00001) (70, 30)};
\end{semilogyaxis}
\end{tikzpicture}
\caption{Beam patterns outputs for two \acp{UE} using QPSK in the multi-user \ac{MIMO} scenario}
\label{fig:mimo-beam}
\end{figure}
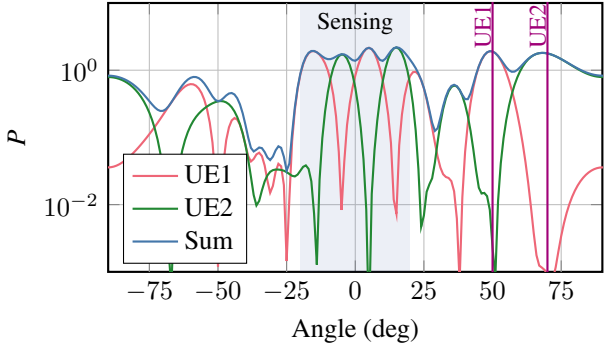

\begin{figure}
	\begin{tikzpicture}
		\begin{semilogyaxis}[
			xlabel=$\text{SNR}_{\text{c}}$ (dB),         ylabel=BER,
			grid=major,
           legend entries= {UE1, UE2},
			legend cell align={left},
			legend pos=south west,
			xmin=5,xmax=30,
			ymin=0.0001,%
			axis line style=thick,
                width=0.95\columnwidth,
                height=0.6\columnwidth,
                legend style={name=leg},
			]
                \addplot[mark=none, color=cb-2, thick] table [x expr= 10*log10(\thisrow{SNR}), y = BERmimo0]
			{figures/mimo/SNRsweep_BER.txt};
                \addplot[mark=none, color=cb-3, thick] table [x expr= 10*log10(\thisrow{SNR}), y = BERmimo1]
			{figures/mimo/SNRsweep_BER.txt};

\end{semilogyaxis}
\end{tikzpicture}
\caption{BER for 2 UEs for different \acp{SNR} using QPSK for modulation and the beam patterns of Fig.~\ref{fig:mimo-beam}}
\label{fig:mimo-ber}
\end{figure}
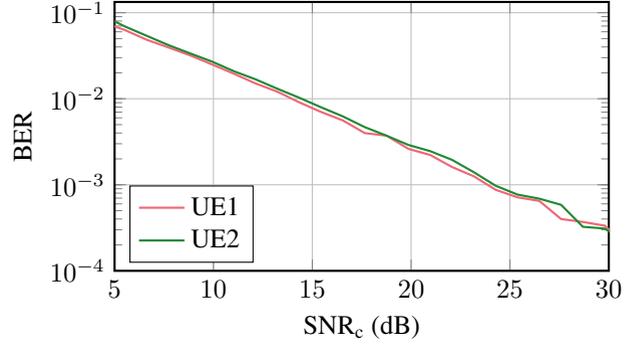

\section{Conclusion}\label{sec:concl}
\acused{ESPRIT}
In this paper, we have proposed a novel loss function for end-to-end trainable \ac{JCAS} systems based on supervised learning. By separating the loss functions from the \ac{SNR}, we have improved training convergence of our system together with its overall performance. We have compared a system with a 16\ac{QAM} constellation to a geometrically shaped constellation. We observed similar behavior in terms of communication and sensing performance, and were able to adjust the trade-off between sensing and communication performance using the trade-off parameter $w_{\text{s}}$. The \ac{QAM} constellation achieved comparable results for object detection and communication compared to the shaped constellation diagrams showing that \ac{JCAS} systems can use legacy constellations. For \ac{AoA} estimation, the shaped constellation achieves lower \acp{MSE} than \ac{QAM}, allowing fine-tuning of the system's performance to achieve different quality of service requirements. The trained object detector and \ac{AoA} estimator both outperform the respective baseline algorithms, namely a Neyman-Pearson-based power detector for object detection and ESPRIT for \ac{AoA} estimation.
By extending our setup, we demonstrate how spatial resources can be used to service multiple \acp{UE}.

In our future work, we want to build on our developed system and proof-of-concept to further explore the effects of multi-user \ac{MIMO} and precoding in \ac{JCAS}. To realize low-complexity solutions, we need to quantify the system complexity in comparison to classical algorithms and explore how reduction of the size of \ac{NN} components affects the performance together with the expected performance-complexity trade-off.
Paradigms such a model-based learning could help to find an appropriate structure of the trainable algorithms.
We can integrate the analysis of this paper with our prior work on multiple target sensing in~\cite{Muth2023}.

%% file: figures/flowgraph_alternative.tex
\tikzset{block/.style={rectangle, thick, draw, minimum width=2cm, minimum height=1cm, rounded corners=1.6mm},font=\small,align=center}
\tikzset{crossing/.style={circle, scale=0.4, fill=black}}
\vspace{0.5cm}
\pgfdeclarelayer{background}
\pgfdeclarelayer{foreground}
\pgfsetlayers{background,main,foreground}
\begin{tikzpicture}[node distance=1.5, text=black, >=latex]
	\node[] (start) {$\vect{m} \in \mathcal{M}^{N_{\text{win}}}$};
        \node[block, node distance=0.5cm, right=of start, fill=KITblue!20] (enc){Modulator};
	
	\node[block, below = of enc, fill=KITblue!20] (beam){Beamformer \\ $K$ antennas};
	\node[node distance=0.5cm, left = of beam] (theta){$\theta_{\min},\theta_{\max},$\\$ \varphi_{\min},\varphi_{\max}$};
	
	\node[draw,circle,inner sep=0, radius=0.4cm,fill=white, thick] (mix) at ($0.5*(beam)+0.5*(enc)+(2,0)$) {\Large{$\otimes$}} ;
	\node[] (mixtext) at (mix) {};  
	\node[block, fill=white] (radch) at ($(mix) + (3.2,1)$) {Sensing channel:\\ Swerling 1 model};
	\node[block, fill=white,node distance=0.7, above=of enc] (ac) {$\text{Corr}(\mat{z_{\text{s}}},\mat{z_{\text{s}}})$};
	\node[block] (commch) at ($(mix) + (3.2,-1)$) {Rayleigh channel};
	\node[crossing] (c1) at ($(mix)+(1,0)$){};
    	
	\node[block, fill=KITblue!20] (detect) at ($(ac) - (3.5,-0.7)$) {Target detection};

	\node[block, fill=KITblue!20] (angle) at ($(ac) + (-3.5,-0.7)$) {Angle estimation};
	\node[crossing] (c2) at ($(ac)-(1.5,0)$){};
	\node[node distance=0.5cm, left = of detect] (llr) {${p}_{\text{T}}$};
	\node[node distance=0.5cm, left = of angle] (phi) {$\hat{{\theta}}$};
	
	\node[block, fill=KITblue!20] (demod) at ($(commch) + (3.5,0)$) {Comm. receiver};
	\node[node distance=0.3cm, below=of commch] (csi)  {CSI $\vect{\gamma}=\nobreak \vect{v}^{\top} \mat{a}_{\text{TX}}(\vect{\varphi})\text{diag}(\vect{\alpha}_{\text{c}})$};
	\node[node distance=0.5cm] (bits) at ($(demod.east)-(-1.5,0)$) {LLRs $\vect{\ell}$\\$ \in \mathbb{R}^{n\times N_{\text{win}}}$};
        \node[] (i1) at ($(angle.east)+(1.4,-0.2)$){$N_{\text{win}}$, $\sigma_{\text{ns}}$};
        \node[] (i2) at ($(detect.east)+(1.4,0.2)$){$N_{\text{win}}$, $\sigma_{\text{ns}}$};
        \node[] (i3) at ($(demod.west)+(-0.75,-0.2)$){$\sigma_{\text{nc}}$};
 
	\draw[thick, ->] (start) -- (enc);
	\draw[thick, ->] (theta) -- (beam);
	\draw[thick, ->] (beam) -| (mix) node[midway,below] {$\vect{v} \in \mathbb{C}^{K}$};
	\draw[thick, ->] (enc) -| (mix) node[midway,above] {$\vect{x} \in \mathbb{C}^{N_{\text{win}}}$};
	\draw[thick, ->] (mix) -- (c1) node[right] {$\mat{y}\in \mathbb{C}^{K\times N_{\text{win}}}$} |-  ($(radch.west)-(0,0.2)$);
	\draw[thick, ->] (mix) -- (c1) |- (commch);
	\draw[thick,->] ($(radch.west)+(0,0.2)$) -- +(-0.915,0) |- (ac) node[midway,right]{$\mat{z}_{\text{s}}$};
	\draw[thick] (ac) -- (c2);
	\draw[thick, ->] (c2) |- ($(detect.east)-(0,0.2)$);
	\draw[thick, ->] (c2) |- ($(angle.east)+(0,0.2)$);
	\draw[thick, ->] (detect) -- (llr);
	\draw[thick, ->] (angle) -- (phi);
        \draw[thick, ->] (i2) -- ($(detect.east)-(0,-0.2)$);
        \draw[thick, ->] (i1) -- ($(angle.east)-(0,0.2)$);
        \draw[thick, ->] (i3) -- ($(demod.west)-(0,0.2)$);
	\draw[thick, ->, dashed] (commch) -- (csi) -| (demod);
	\draw[thick, ->] ($(commch.east)+(0,0.2)$) -- ($(demod.west)+(0,0.2)$) node[midway,above]{${\vect{z}}_{\text{c}}$};
	\draw[thick, ->] ($(demod.east)-(0,0)$) -- (bits);
	\draw[thick, dashed] (beam.south) |- (csi);

 \begin{pgfonlayer}{background}
\node[draw, minimum size=2cm, rectangle,minimum width = 7.8cm, minimum height=6.5cm, KITgreen, fill=KITgreen!20,thick, dotted,
	label=south:{Base station}] (A) at ($(ac)+(-1,-1.99)$) {};
 \pic[ node distance=0.5cm,left = of theta, scale=1, gray, very thick] (bs) {MBS};
 \pic[ scale=0.4, KITblue, very thick] (car) at ($(radch.north east)+(-1.5,-0.35)$) {car};
 \node[label=north:{Sensing target}] (labcar) at ($(radch.north east)+(0.2,0.6)$) {};
 \node[draw, minimum size=2cm, rectangle,minimum width = 4.8cm, minimum height=2.5cm, KITgreen, fill=KITgreen!20,thick, dotted,
	label=south:{User equipment}] (A) at ($(demod)+(1,0)$) {};
 \pic[thick, fill=white, scale=0.25] (UE1) at ($(demod)+(0.8,0.15)$) {phone};
\end{pgfonlayer}
\end{tikzpicture}

%% file: tex_files/attachment.tex
\section{\ac{CRB} Derivation}\label{app:crb}
We know from \cite[Ch. 8.4]{Trees2002} that the \ac{CRB} for phase recovery of phase $\gamma=\pi \sin(\theta)$ is given by
\begin{align}
    {C}_{\text{CR}} (\gamma) = \frac{\sigma_{\text{n}}^2}{2N} \Re \left\{  \left[ \mat{S}_{\text{f}}\left[\left(\mat{I}+\mat{A}^H \mat{A} \frac{\mat{
    S}_{\text{f}}}{\sigma_{\text{n}}^2}\right)^{-1} \right.\right.\right. \notag \\ 
    \left.\left.\left. \left(\mat{A}^H \mat{A} \frac{\mat{
    S}_{\text{f}}}{\sigma_{\text{n}}^2} \right)\right]\right] \odot \mat{H}^\top \right\}^{-1}. \label{eq:CRB:general}
\end{align}
In \eqref{eq:CRB:general}, $N$ denotes the number of samples and $\sigma_{\text{n}}^2$ is the noise power. For the single source case, the signal spectral matrix $\mat{S}_{\text{f}}$ reduces to a scalar $s_{\text{f}} = \beta \sigma_{\text{s}}^2$,
with transmit beamforming gain $\beta$ and transmit power $\sigma_{\text{s}}$.
The Matrix $\mat{H}$ is defined as
\begin{align}
    \mat{H} = \dot{\mat{A}}_{\gamma}^H \left[ \mat{I}- \mat{A}(\mat{A}^H\mat{A})^{-1} \mat{A}^H  \right] \dot{\mat{A}_{\gamma}},
\end{align}
with $\dot{\mat{A}}_{\gamma}$ denoting the derivative of $\mat{A}$ with respect to $\gamma$.
The steering matrix $\mat{A}$ simplifies for a uniform linear array to a steering vector
 \begin{align}
    \vect{a}&=[1,\exp(-\mathrm{j} \gamma), \exp(-\mathrm{j} 2 \gamma), \ldots, \exp(-\mathrm{j}(K-1) \gamma)]^\top
\end{align}
and the derivative $\dot{\vect{a}}_{\gamma}$ with respect to $\gamma$ to
\begin{align}
    \dot{\vect{a}}_{\gamma} &= [1, -\mathrm{j} \exp(-\mathrm{j} \gamma),  \ldots, -\mathrm{j} (K-1) \exp(-\mathrm{j} (K-1)\gamma)]^\top.
 \end{align}
For the CRB concerning the angle of arrival $\theta$ with $\gamma$ being a function of $\theta$, we can rewrite
\begin{align}
    \mat{C}_{\text{CR}} (\gamma) = \dot{\mat{F}}^\top \mat{C}_{\text{CR}} (\theta) \dot{\mat{F}}, \quad (\dot{\mat{F}})_{i,j} = \frac{d \gamma}{d \theta}.
\end{align}
Evaluating the inner terms of~\eqref{eq:CRB:general} for one source/target yields:
\begin{align}
    \left(\mat{I}+\mat{A}^H \mat{A} \frac{\mat{S}_{\text{f}}}{\sigma_{\text{n}}^2}\right)^{-1} 
    &= \frac{\sigma_{\text{n}}^2}{\sigma_{\text{n}}^2 + K\beta \sigma_{\text{s}}^2},\\
    \left(\mat{A}^H \mat{A} \frac{\mat{
    S}_{\text{f}}}{\sigma_{\text{n}}^2} \right) &= \frac{K \beta \sigma_{\text{s}}^2}{\sigma_{\text{n}}^2}.
\end{align}
Now we calculate $\mat{H}$, which is also a scalar for the single source case:
\begin{align}
    H &= \dot{\vect{a}}_{\gamma}^H \left[ \vect{I}- \vect{a}(\vect{a}^H\vect{a})^{-1} \vect{a}^H  \right] \dot{\vect{a}_{\gamma}}\\
    &= \dot{\vect{a}}_{\gamma}^H \dot{\vect{a}}_{\gamma} - \dot{\vect{a}}_{\gamma}^H \vect{a}(\vect{a}^H\vect{a})^{-1} \vect{a}^H \dot{\vect{a}}_{\gamma} \\
    &= \sum_{k=0}^{K-1} k^2 - \left(\sum_{k=0}^{K-1}k \right) \cdot \frac 1K \cdot \left(\sum_{k=0}^{K-1}k \right)\\
    &= \frac{K(K-1)(2K-1)}{6} - \frac{K(K-1)^2}{4} \\
    &= \frac{K^3 - K}{12} = \frac{K(K^2-1)}{12}.
\end{align}
We can formulate the CRB concerning $\gamma$:
\begin{align}
    \mat{C}_{\text{CR}} (\gamma) &= \frac{\sigma_{\text{n}}^2}{2N} \Re \left\{  \left[ \beta \sigma_{\text{s}}^2 \left[\frac{\sigma_{\text{n}}^2}{\sigma_{\text{n}}^2 + K\beta \sigma_{\text{s}}^2}  \frac{K \beta \sigma_{\text{s}}^2}{\sigma_{\text{n}}^2} \right]\right] \cdot \notag\right.\\ 
    & \left. \frac{K^3 - K}{12} \right\}^{-1}.
\end{align}
Calculating the derivative of $\gamma$ with respect to $\theta$ yields
\begin{align}
    \frac{\mathrm{d} \gamma}{\mathrm{d} \theta} &= \frac{\mathrm{d} \pi \sin(\theta)}{\mathrm{d} \theta} \\
    &= \pi \cos(\theta).
\end{align}
Putting everything together yields \eqref{eq:CRB}.

\section{Sufficient Statistic}\label{app:suffstat}
In this section, we reason the use of correlation preprocessing in the sensing receiver by proving that it is a sufficient statistic for object detection and \ac{AoA} estimation. A sufficient statistic is a set of values that contains all information of a set of measurement samples concerning the estimation of a certain parameter, i.e., an optimal estimator based on the sufficient statistic achieves the same estimation accuracy as an optimal estimator based on the measurement samples.

For \ac{AoA} estimation, we can show the sufficiency mathematically following~\cite{Bekkerman2006}. The log-likelihood function to estimate the \ac{AoA} $\theta$ of the measurement samples $\mat{Z}_{\text{s}}$ can be written in an \acs{AWGN} scenario as:
\begin{align}
    \log & f_{\mat{Z}_{\text{s}}}(\mat{Z}_{\text{s}};\theta) = -KN_{\text{win}}\log(\pi \sigma_{\text{ns}}^2) - \frac{1}{\sigma_{\text{ns}}^2} \sum_{n=1}^{N_{\text{win}}} ||\vect{z}_{\text{s}}||^2 \notag\\
    &+ \frac{2\sqrt{N_{\text{win}}}}{\sigma_{\text{ns}}^2} \Re \left\{ \alpha_{\text{s}} \left(\vect{a}_{\text{RX}}({\theta}) \vect{a}_{\text{TX}}({\theta})^\top\right)^H \mat{Z}_{\text{s}} \mat{y}^H  \right\} \notag\\
    &- \frac{|\alpha_{\text{s}}|^2N_{\text{win}}}{\sigma_{\text{ns}}^2} \vect{a}_{\text{RX}}^H({\theta}) \mathrm{Corr}(\mat{Y},\mat{Y})^H \vect{a}_{\text{TX}}({\theta}) \vect{a}_{\text{RX}}^H({\theta}) \vect{a}_{\text{TX}}({\theta})
\end{align}
A sufficient statistic $\mat{\eta}$ can be found through the factorization theorem, i.e., by rewriting the log-likelihood as $\log f_{\mat{Z}_{\text{s}}}(\mat{Z}_{\text{s}};\theta)= f_1(\mat{Z}_{\text{s}}) + f_2(\theta,\mat{\eta})$. According to~\cite{Bekkerman2006}, the sufficient statistic can be extracted as
\begin{align}
    \mat{\eta} = \frac{1}{\sqrt{N_{\text{win}}}} \mat{Z}_{\text{s}} \mat{y}^H. \label{eq:suff-stat}
\end{align}
In case $\mat{Y}$ is unknown, $\mat{\eta}$ can be approximated with the projection of the received signal on the signal space as in~\cite{Ziskind1988}:
\begin{align}
    \mat{\eta} &\approx \frac{1}{\sqrt{N_{\text{win}}}} \mat{Z}_{\text{s}} \left( \left(\vect{a}_{\text{RX}}({\theta}) \vect{a}_{\text{TX}}({\theta})^\top \right)^{\dagger}  \mat{Z}_{\text{s}}\right)^H \\
    &= \frac{1}{\sqrt{N_{\text{win}}}} \mat{Z}_{\text{s}}  \mat{Z}_{\text{s}}^H \left(\left(\vect{a}_{\text{RX}}({\theta}) \vect{a}_{\text{TX}}({\theta})^\top \right)^{\dagger}\right)^H   ,\label{eq:suff-approx}
\end{align}
with $(\cdot)^{\dagger}$ denoting the Moore-Penrose pseudoinverse.
In our case, the transmit signal is known, but using the approximation \eqref{eq:suff-approx} allows extension to bistatic scenarios, resulting in the approximated sufficient statistic
\begin{align}
    \Tilde{\mat{\eta}} = \mathrm{Corr} (\mat{Z}_{\text{s}},\mat{Z}_{\text{s}}). \label{eq:suff-corr}
\end{align}
Notably, the ESPRIT algorithm that we use as a baseline uses \eqref{eq:suff-corr} as input. The comparison of the proposed algorithm and the baseline becomes more fair if both systems estimate based on the same statistic.

For object detection, the total power of the received signal projected on the \acp{AoA} is a sufficient statistic as shown in ~\cite{Bekkerman2006} and can be calculated from the correlation matrix in \eqref{eq:suff-stat}. Under an unknown transmit signal, the approximation in \eqref{eq:suff-approx} applies. Therefore, we can feed \eqref{eq:suff-corr} into the detection \ac{NN} and approach optimal performance after training. The baseline \ac{NP}-detector is also only based on the total power of the signal and does not rely on a known transmit signal.